\documentclass[final,3p,8pt]{elsarticle}

\usepackage{amssymb}
\usepackage{array}
\usepackage{mathrsfs}
\usepackage{amscd}
\usepackage{amsmath}
\usepackage{amsfonts}
\usepackage{amsthm}
\usepackage{psfrag}
\usepackage{textcomp}
\usepackage{bm}
\usepackage{dsfont}
\usepackage{algorithm}
\usepackage{algpseudocode}

\usepackage{amsthm}

 \usepackage{lineno}

\journal{Proceedings of the Royal Society A: Mathematical, Physical and Engineering Sciences}

\usepackage{amssymb}

\biboptions{square,sort&compress}

\usepackage[figuresright]{rotating}
\usepackage{moreverb}
\usepackage{amssymb}
\usepackage{array}
\usepackage{mathrsfs}
\usepackage{amscd}
\usepackage{amsmath}
\usepackage{amsfonts}
\usepackage{amsthm}
\usepackage{psfrag}
\usepackage{textcomp}
\usepackage{bm}
\usepackage{color}
\usepackage{float}
\usepackage[latin1]{inputenc} 	
\usepackage{afterpage}
\usepackage{tabularx}
\usepackage{booktabs}
\usepackage{romannum}
\usepackage{multirow}
\usepackage[normalem]{ulem} 
\usepackage{cancel}
\usepackage{soul}
\usepackage{cancel}

\begin{document}


\begin{frontmatter}





\title{Dispersion relations of generalized one-dimensional phononic crystals}

\author[upv2,icl]{Mario L\'azaro\corref{cor}}
\ead{malana@upv.es}
\author[upv2]{Vicent Romero-Garc\'ia}
\author[icl]{Richard Wiltshaw}
\author[icl]{Richard V. Craster}

\cortext[cor]{Corresponding author. Tel +34 963877000 (Ext. 76732). Fax +34 963877189}
\address[upv2]{Instituto Universitario de Matem\'atica  Pura y Aplicada, 
	Universitat Polit\`ecnica de Val\`encia, 46022 (Spain)}
\address[icl]{Department of Mathematics, Imperial College London, London, SW7 2AZ, UK}

\begin{abstract}

We present a comprehensive method for determining {both exact and approximate} dispersion {relations} for one-dimensional {resonant phononic} crystals, applicable to a wide range of structures, regardless of their specific characteristics. This general framework employs a unified mathematical model, referred to as generalized  {one-dimensional (1D) phononic crystal}, in which {different} types of {waves} and scatterers{/resonators} {can be} {considered} by adjusting certain parameters. The generalized {1D phononic crystal} consists of both a host {one-dimensional} homogeneous elastic {material} with physical properties represented in matrix form and an arbitrary set of scatterers {within the unit cell,} including resonators (discrete and continuous), small material inclusions, or variations in cross-sectional area. Based on general assumptions, {and imposing the periodicity and Bloch solutions} we develop a matrix-based algorithm utilizing the plane wave expansion method to derive the solution. Additionally, we propose an iterative procedure that provides analytical expressions for the first- and second-order terms, particularly useful in the context of weak scattering. The convergence conditions of the method are rigorously defined. The {efficiency of the} approach is demonstrated through several numerical examples, highlighting its versatility in different waveguide configurations and scattering scenarios.

\end{abstract}


\end{frontmatter}


\section{Introduction}



In recent decades, phononic crystals with resonant characteristics, often referred to as phononic metamaterials, have attracted considerable attention due to their capability to manipulate low-frequency wave propagation. This control is achieved through periodic media incorporating resonator arrays that can be tuned to either specific frequencies \cite{WangG-2005} or a range of frequencies \cite{Miranda-2019, WangZ-2013, Xiao-2012c, Xiao-2013, Wiltshaw-2023}.
Metamaterials have been broadly embraced in multiple areas of wave physics as a means to engineer subwavelength-scale devices. This is achieved by harnessing resonance effects, with notable applications in fields such as photonics \cite{Monticone-2017, Ali-2022} and phononics \cite{Cummer-2026, Brule-2022, Muhammad-2020, Krushynska-2023}. Applications span areas such as energy harvesting \cite{Chen-2014, Chaplainc-2020, DePonti-2020, DePonti-2021} and seismic wave mitigation systems \cite{Colombi-2016a, Colombi-2016b, Lott-2020, Mu-2020, Yakovleva-2021}. The analysis of heterogeneous media and elastic solids with strong material contrast, particularly those exhibiting natural resonances and modified characteristics under perturbation, holds significant importance in experimental modal testing and structural diagnostics \cite{Croxford-2007}. To investigate the vibrational response of such systems, analytical models supported by finite element simulations have been employed to examine wave scattering and energy transmission in the presence of multiple resonating inclusions \cite{WuJ-1998, Brennan-1999, Mace-2007, TanC-1998}, as well as to understand the dynamic response of elastic waveguides featuring spatially varying mechanical properties \cite{LeeJ-2016, Adamek-2015, Aya-2012, ZhangK-2017, WangP-2015, Wiltshaw-2022}. Configurations incorporating Rayleigh-type beams have been proposed to achieve control of wave propagation through the coupling between axial and bending modes \cite{Movchan-2022a, Movchan-2022b, Movchan-2023}. In the context of flexural wave propagation in thin plates, arrays of localized resonators require the use of plane wave expansion techniques in conjunction with multiple scattering theory \cite{Wiltshaw-2023, Torrent-2012, Wiltshaw-2020, Movchan-2017a, Movchan-2018, Ruzzene-2017}.\\

A specific class of locally resonant elastic metamaterials, especially applicable to one-dimensional (1D) or two-dimensional (2D) wave propagation, incorporates local resonators as elastic substructures. Examples include elastic rings around a rod \cite{Hussein-2012a, Xiao-2012b} or beam \cite{Yu-2006} and  pillar-type resonators arranged on an elastic plate \cite{Pennec-2008, WuT-2008, WuT-2011, Xiao-2012a, Oudich-2010a, Hussein-2013c, ZhangS-2013}. The discrete lumped parameter model has emerged as a core model for 1D wave propagation in acoustic or elastic metamaterials \cite{HuangH-2009,Hussein-2021,WangX-2024}. Local resonances from attached substructures have been applied in sound isolation \cite{Oudich-2016}, elastic waveguiding \cite{Pennec-2008, Oudich-2010b},  lensing design \cite{Bigoni-2013}, and topological insulation \cite{Jin-2018, ChenJ-2017}. A recent review covers advances in elastic metamaterials and metasurfaces that exhibit surface vibrational resonances \cite{Jin-2021}. Additionally, for the mathematical modeling of impedance contrast due to cracks, internal elastic springs in conjunction with spectral element method \cite{Krawczuk-1994a,Krawczuk-1996,Krawczuk-2003b,Krawczuk-2002a,Krawczuk-2004b} or custom-designed finite elements \cite{Yeung-2019} have been developed. In the realm of nanomaterial mechanics, theories relying on material distinctions have been suggested for analyzing how structures with numerous cracks spread across their length behave and how they affect the modal characteristics. This applies to both flexural waves (in nanobeams) \cite{Loghmani-2018a} and longitudinal waves (in nanorods) \cite{Loghmani-2018b}.\\

Traditionally, analytical procedures for obtaining the dispersion relation in one-dimensional homogeneous media have been presented separately in the literature \cite{Doyle-1997,Graff-1999,Royer-2022}, starting with classical media such as the bar (longitudinal and torsional waves) and the Euler-Bernoulli beam (flexural waves) to continue with modified models (higher-order rod or beam theories). The intricacy grows as the number of variables increases, rendering the analytical solutions for the differential equation of time-varying motion difficult to handle, particularly when factoring in localized disruptions within the medium like attached structures or changes in properties of cross sections or material. Hence, it's beneficial to present the constitutive equations in a first-order format \cite{Lazaro-2025a}, as it simplifies the expression of continuity conditions across progressively expanding degrees of freedom (such as displacement, rotation, force, moments, etc.) into a matrix. 
\\

This work introduces an analytical approach for deriving the dispersion relation in one-dimensional phononic crystals with any number of embedded scatterers, which can take the form of point resonators, small material inclusions, and/or changes in cross-sectional area. The model to reproduce small inclusions as point perturbations, introduced in ref. \cite{Lazaro-2025a} in a context of multiple scattering, presents the advantage of following the same scheme used for generic resonators, which allows us to incorporate it into this work in the context of periodic phononic structures. Building on this, we introduce a unified mathematical framework that applies to any type of scatterer, enabling a generalized approach. This mathematical model is called a {\em generalized 1D phononic crystal} to highlight the fact that a large number of both elastic structures and scatterers can be considered under the same mathematical framework. The problem is addressed using the plane wave expansion technique, and an iterative procedure is developed to express the dispersion relation as a perturbation of the modes. This approach, based on the weak scattering assumption, provides analytical solutions that can help to predict dispersion and propagation. We also rigorously establish the numerical convergence criteria for the iterative method. The theoretical findings are validated through three numerical examples: (i) an Euler-Bernoulli beam with point spring-mass resonators, (ii) a Timoshenko beam with continuous resonators (embedded beams), and (iii) waveguides with small inclusions affecting both longitudinal and flexural waves. The latter has been developed in the supplementary material to the article.

\section{The generalized 1D phononic crystal}

In this work, we are mainly interested in the dispersion relations of an elastic one-dimensional phononic crystal made of an arrangement of \(N\) perturbations or scatterers distributed along each unit cell of length \(L\). Before outlining the methodology proposed in this paper, it is important to first present the basis of the analytical model used to describe the homogeneous elastic host  medium, i.e. without any perturbations. This background will provide essential context for understanding the effects of scatterers on wave propagation in the 1D phononic crystal.\\

One-dimensional models enable us to represent the displacement pattern of any cross-section using reference degrees of freedom that solely rely on the longitudinal position $x$. By employing Hamilton's principle in conjunction with cross-sectional kinematic assumptions, a series of partial differential equations emerge in space-time dimensions $(x,t)$, expressed in terms of $m$ kinematic parameters and $m$ generalized forces, denoted respectively as $\bm{v}(x,t)$ and $\bm{F}(x,t)$. Generally, all system degrees of freedom can be encapsulated into a column vector, referred to as the state-vector $\bm{u}(x,t)$ of dimension $2m$, structured as follows:
\begin{equation}
	\bm{u}(x,t) = \left\{\begin{array}{cc} \bm{v}(x,t)  \\ \bm{F}(x,t) \end{array} \right\} .
	\label{eq000}
\end{equation}
Assuming that the 1D elastic host medium is homogeneous and considering harmonic response of the type $\bm{u}(x,t) = \mathbf{u}(x) e^{i \omega t}$ where $\omega$ is the angular frequency, the frequency domain differential equations of motion, governing the system can be expressed as
\begin{equation}
	\frac{\textrm{d} \mathbf{u}}{\textrm{d} x} = \mathbf{A} \, \mathbf{u} + \mathbf{q}(x) \, ,
	\label{eq001}
\end{equation}
The $\mathbf{A}$ matrix, sized $2m \times 2m$, encapsulates parameters which depend on frequency, reflecting the medium's stiffness, mass distribution, and rotational inertia properties. The vector $\mathbf{q}(x)$ models the spatially varying external forces applied to the system, with nonzero components confined exclusively to its lower half, corresponding to generalized force terms. To better understand the practical implications, Table \ref{tab01} presents explicit forms of the state vector, loading vector, and system matrix $\mathbf{A}$ for a variety of representative cases. These include traditional and advanced formulations describing wave propagation in longitudinal, bending, and torsional modes, along with a case study demonstrating coupling effects between flexural and torsional vibrations.\\

\begin{table}[h]	
	\begin{center}
		\scriptsize{
\begin{tabular}{lccccr}
	Model 				   		&			$\mathbf{u}(x)$	&			$\mathbf{q}(x)$ & Matrix $\mathbf{A}$	 	\\ 
	\hline \\
	\begin{tabular}{l} Longitudinal \\ (classical rod) \\ $2m = 2$  	\end{tabular}	 
								 &			 $\left\{\begin{array}{cc} u \\ N_x \end{array} \right\}$
								 &			 $\left\{\begin{array}{cc} 0 \\ -q_x \end{array}  \right\}$ 
								 &			 $\left[ \begin{array}{cc} 0  & 1/EA \\ - \varrho A \omega^2 & 0 \end{array} \right]$  \\ \\
	\begin{tabular}{l} Longitudinal \\ (Love's rod)  	\\ $2m=2$ \end{tabular}	 
								&			 $\left\{\begin{array}{cc} u \\ N_x \end{array} \right\}$
								&			 $\left\{\begin{array}{cc} 0 \\ -q_x \end{array}  \right\}$ 
								&			 $\left[ \begin{array}{cc} 0  & 1/(EA-\varrho I_x \nu^2 \omega^2) \\ - \varrho A \omega^2 & 0 \end{array} \right]$  \\ \\
	\begin{tabular}{l} Torsional  \\ (Sain--Venant) \\ $2m = 2$  	\end{tabular}	 
								&			 $\left\{\begin{array}{cc} \theta_x  \\ T_x \end{array} \right\}$
								&			 $\left\{\begin{array}{cc} 0 \\ -m_x \end{array}  \right\}$ 
								&			 $\left[ \begin{array}{cc} 0  & 1/GJ \\ - \varrho I_x \omega^2 & 0 \end{array} \right]$  \\ \\		
	\begin{tabular}{l} Torsional  \\ (Open thin-walled \\ Vlassov beams) \\ $2m = 4$ 	\end{tabular} 	 
								&			 $\left\{\begin{array}{cc} \theta_x \\ \varphi \\ T_x \\  M_w \end{array} \right\}$
								&			 $\left\{\begin{array}{cc} 0 \\ 0 \\ -m_x  \\ 0 \end{array}  \right\}$ 								
								&			 $\left[ \begin{array}{cccc} 0  & 1 &  0 & 0 \\ 0 & 0 & 0 & 1/EI_w  \\ 
									- \varrho I_x \omega^2 & 0 & 0 & 0 \\ 0 &  \varrho I_w \omega^2 - GJ  & -1 & 0  \end{array} \right]$ 		\\ \\														
	\begin{tabular}{l} Flexural \\ (Euler-Bernoulli) \\  	$2m = 4$ \end{tabular} 	 
								&			 $\left\{\begin{array}{cc} w \\ \theta_y \\ V_z \\  M_y \end{array} \right\}$
								&			 $\left\{\begin{array}{cc} 0 \\ 0 \\ -q_z \\ - m_y \end{array}  \right\}$ 
								&			 $\left[ \begin{array}{cccc} 0  & 1 &  0 & 0 \\ 0 & 0 & 0 & 1/EI_y  \\ 
																		- \varrho A \omega^2 & 0 & 0 & 0 \\ 0 &  0  & -1 & 0  \end{array} \right]$ 	\\	\\			
	\begin{tabular}{l} Flexural \\ (Timoshenko) \\  	$2m = 4$ \end{tabular} 	 
								&			 $\left\{\begin{array}{cc} w \\ \theta_y \\ V_z \\  M_y \end{array} \right\}$
								&			 $\left\{\begin{array}{cc} 0 \\ 0 \\ -q_z \\ - m_y \end{array}  \right\}$ 
								&			 $\left[ \begin{array}{cccc} 0  & 1 &  1/GA_z & 0 \\ 0 & 0 & 0 & 1/EI_y  \\ 
									- \varrho A \omega^2 & 0 & 0 & 0 \\ 0 &  -\varrho I_y \omega^2  & -1 & 0  \end{array} \right]$ 	\\	\\																							
	\begin{tabular}{l} Flexural-torsional \\ $y$--symmetry \\ (Timoshenko, \\ Saint-Venant) \\ $2m = 6$	\end{tabular} 	 
								&			 $\left\{\begin{array}{cc} w \\ \theta_y \\ \theta_x \\ V_z \\ M_y \\  T_x \end{array} \right\}$
								&			 $\left\{\begin{array}{cc} 0 \\ 0 \\ 0 \\ -q_z \\ - m_y \\ - m_x  \end{array}  \right\}$ 
								&			 $\left[ \begin{array}{cccccc} 0  & 1 &  0 & 1/GA_z & 0 & 0 \\ 0 & 0 & 0 &  0 &  1/EI_y & 0  \\ 
																			0 & 0 & 0 & 0 & 0 & 1 / GJ \\
																		- \varrho A \omega^2 & 0 & - \varrho A y_G \omega^2 & 0 & 0 & 0 \\ 
																			0 &  - \varrho I_y \omega^2 & 0   & -1 & 0 & 0 \\
																			- \varrho A y_G \omega^2 & 0 &   - \varrho I_x \omega^2 & 0 & 0 & 0 \end{array} \right]$ 					\\ \\
	\begin{tabular}{l} Flexural-longitudinal \\  Rod-Beam model \\ $2m = 6$	\end{tabular} 	 
								&			 $\left\{\begin{array}{cc} u \\ w \\ \theta_y \\ N_x \\ V_z \\ M_y \end{array} \right\}$
								&			 $\left\{\begin{array}{cc} 0 \\ 0 \\ 0 \\ -q_x \\ - q_z \\ - m_y  \end{array}  \right\}$ 
								&			 $\left[ \begin{array}{cccccc} 
																		0  & 0 &  0 & 1/EA & 0 & 0 \\ 
																		0 & 0 & 0 &  0 &  1/GA_z & 0  \\ 
																		0 & 0 & 0 & 0 & 0 & 1 / EI_y \\
																	- \varrho A \omega^2 & 0 &  0 & 0 & 0 & 0 \\ 
																			0 &  - \varrho A \omega^2 & 0   & 0 & 0 & 0 \\
																		0 & 0 &   - \varrho I_y \omega^2 & 0 & 0 & 0 \end{array} \right]$ 					\\ \\
																		 \hline 																								
\end{tabular}}
		\caption{The generalized beam model in the absence of scatterers. Specification of the state vector $\mathbf{u}(x)$, load vector $\mathbf{q}(x)$, and system parameter matrix $\mathbf{A}$ for various elastic waveguide types, including those supporting purely longitudinal, torsional, and flexural wave motions, as well as cases involving coupled wave phenomena.}
		\label{tab01}
	\end{center}	
\end{table}

A particularly illustrative demonstration of this approach appears when  comparing the Euler-Bernoulli and Timoshenko beam models (refer to entries 5 and 6 in Table \ref{tab01}). Typically, the Timoshenko model is  as a system of second-order partial differential equations concerning displacement and rotation, contrasting with the Euler-Bernoulli model, which is a classical fourth-order equation focusing on displacement exclusively. However, in this context, we will adopt a unified approach for both models, augmenting the matrix $\mathbf{A}$ with additional terms to incorporate considerations for shear deformation and rotational inertia.\\

In the development of our approach the dispersion relation of the host medium without scatterers, which represents the unperturbed dispersion relation is of interest. Considering Eq.~\eqref{eq001} in absence of external loads $\mathbf{q}(x)\equiv \mathbf{0}$, we can check for solutions of the form $\mathbf{u}(x) = \hat{\mathbf{u}} \, e^{i k x}$ leading to the following eigenvalue problem
\begin{equation}
	\left[\mathbf{A} - i k \, \mathbf{I} \right] \, \hat{\mathbf{u}} = \mathbf{0} \ , 
	\label{eq164}
\end{equation}
where $\mathbf{I}$ stands for the identity matrix of size $2m$ and $k$ represents the wavenumber. Any waveguide governed by the matrix $\mathbf{A}$ with state--vector $\mathbf{u}(x)$ will present in general $2m$ modes, with $m$ modes corresponding to rightwards waves ($k < 0$) and $m$ modes corresponding to leftwards waves ($k > 0$). These modes can be either propagating or evanescent depending on the complex nature of wavenumber $k$. For instance, longitudinal waves in rods have $m=1$ propagating modes in each direction. Low frequency flexural waves present $2m=4$ modes, one propagating and one evanescent for both rightward and leftward waves. High frequency beam waves (Timoshenko beam) present $m=2$ propagating modes at each direction (bending and shear waves). Denoting $\mathbf{u}_j$ and $\mathbf{v}_j$ to be the right and left eigenvectors associated to each mode with eigenvalue $i k_j$, i.e. for the matrix $\textbf{A}$ governing the conserved quantities along the empty 1D elastic medium

For any waveguide characterized by the matrix $\mathbf{A}$ and the state vector $\mathbf{u}(x)$, there generally exist $2m$ distinct wave modes. Among these, $m$ correspond to waves traveling in the positive $x$-direction (rightwards, with $k < 0$) and $m$ correspond to waves propagating in the negative $x$-direction (leftwards, with $k > 0$). Depending on whether the wavenumber $k$ is real or complex, these modes may be either traveling or evanescent. As an example, longitudinal waves in rods feature a single propagating mode ($m=1$) in each direction. Low-frequency flexural waves, on the other hand, exhibit four modes ($2m=4$) comprising one propagating and one evanescent wave for both directions. At higher frequencies, Timoshenko beam waves display two propagating modes per direction ($m=2$), corresponding to bending and shear motions. Let $\mathbf{u}_j$ and $\mathbf{v}_j$ denote the right and left eigenvectors associated with the $j$-th mode, having eigenvalue $i k_j$, where $\mathbf{A}$ is characteristic of the empty one-dimensional elastic medium.
\begin{equation}
	\mathbf{A} \mathbf{u}_j = i k_j \, \mathbf{u}_j \ , \quad 
	\mathbf{v}_j^T \, \mathbf{A}  = i k_j \, \mathbf{v}_j^T \ , \quad 	
	\mathbf{v}_j^T \, \mathbf{u}_l  = \delta_{jl} \, , \quad 1 \leq j,l \leq 2m \ , 
	\label{eq066}
\end{equation}
where $\delta_{jl} $ stands for the Kronecker--delta coefficient. \\

\begin{figure}[h]%
	\begin{center}
		\begin{tabular}{c}
			\includegraphics[width=12cm]{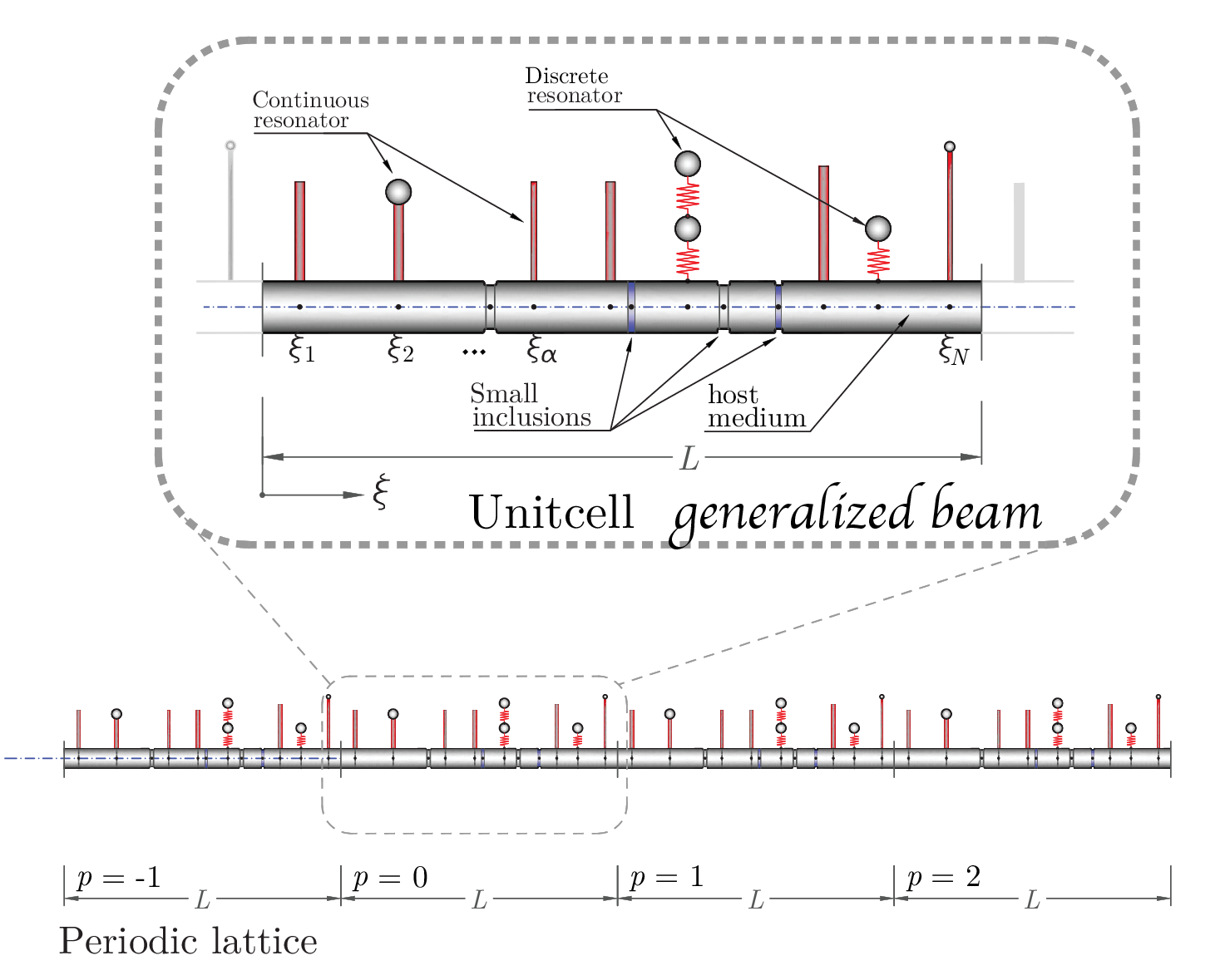} 
		\end{tabular}			
		\caption{Generalized 1D phononic crystal formed by a elastic host medium with an arbitrary number $N$ of scatterers distributed in the unit cell of length $L$. The 1D structure has properties arranged in matrix $\mathbf{A}$, see Table \ref{tab01}. The scatterers considered can be of two types: (1) linear resonators attached pointwise to the waveguide and (2) small-width inclusions with other material or changes in cross-sectional dimensions. The properties of scatterers are introduced in matrices $\mathbf{K}_\alpha$, see Table \ref{tab02}}%
		\label{fig01}%
	\end{center}
\end{figure}

{Now we analyze the case in which scatterers are present in the 1D elastic medium, i.e., considering Eq.~\eqref{eq001} with external loads $\mathbf{q}(x)\neq \mathbf{0}$}. For this work, we define a unit cell of length $L${, so covering from $x=0$ to $x=L$} along the {1D elastic medium, in which} $N$ scatterers {are placed} inside at different and arbitrary positions $\xi_1,\ldots,\xi_N$  (see Fig \ref{fig01}). The positions of the scatterers can be expressed as
\begin{equation}
	x_{p \alpha} = p \, L + \xi_\alpha \ , \qquad p = 0, \pm 1, \pm 2, \ldots \ , \qquad 1 \leq \alpha \leq N \ ,
	\label{eq108}
\end{equation}
{where the subindex $p$ represents the unit cell and $\alpha$ the scatterer identification.} We will {consider}  two types of scatterers:
\begin{description}
	\item [Resonators:] Either discrete or continuous elastic attached substructures can be considered, depending on whether they have a finite or infinite number of internal resonance frequencies, respectively. The resonator vibrates according to the internal resonances inducing forces back through the attachment point, radiating the effect to the rest of the host medium. Mathematically, each resonator is characterized by a frequency-dependent transfer matrix $\mathbf{T}^{(\alpha)}$ locally-defined  along its length, which relates generalized degrees of freedom and forces between the attaching point and the free end. For more details about this matrix, the reader is refereed to Sec. 1 from the supplementary material. 
	\item [Inclusions:] We define inclusions as those regions, defined along the 1D elastic medium and of relatively small size with respect to the wavelength, where the material and/or section properties change. These perturbations of the beam properties  induce changes in the transmission pattern and generate reflections. We consider that each inclusion is homogeneous along its width $\Delta x_\alpha$ and inside the waves are travelling according to Eq. \eqref{eq001} but with another different properties, say $\mathbf{A}_\alpha$, different to $\mathbf{A}$, generating a perturbation. 
\end{description}

In both cases, the perturbations induce certain radiating forces located at the scatterers, giving rise to the scattered field. Thus, the total wavefield can be obtained introducing certain external concentrated forces as located at the scatterers with the form
\begin{equation}
	\mathbf{q}(x) =\sum_{p} \sum_{\alpha=1}^N \mathbf{K}_\alpha \, \mathbf{u}(x_{p \alpha}) \, \delta(x - x_{p \alpha}) 
	\label{eq165}
\end{equation}
where $\delta(x)$ stands for the Dirac-delta function and $\mathbf{u}(x_{p\alpha})$ is the  state-vector evaluated at the $\alpha$th scatterer at the $p$th unit cell. The $2m \times 2m $--matrix $\mathbf{K}_\alpha $ is characteristic of each scatterer and depends on whether it is a resonator-type or a inclusion-type, showing a completely different form in both cases. The $\mathbf{K}_\alpha$  matrix has been derived for resonator-type and inclusion-type scatterers in references \cite{Lazaro-2025a,Lazaro-2024c}, respectively. In Secs. 1 and 2 included in the supplementary material, some details on the derivations can be found, useful from a practical point of view.  Table \ref{tab02} summarizes these results.\\

The combination of the host-medium, whose properties are contained within matrix $\mathbf{A}$, and the $N$ arbitrarily distributed scatterers characterized by the set of $N$ matrices $\mathbf{K}_\alpha$,  constitute together the new proposed concept called {\em generalized 1D phononic crystal}. Unifying Eqs. \eqref{eq001} and \eqref{eq165}, it obeys the following differential equation 
\begin{equation}
	\frac{\textrm{d} \mathbf{u}}{\textrm{d} x} = \mathbf{A} \, \mathbf{u}  + \sum_{p} \sum_{\alpha=1}^N \mathbf{K}_\alpha \, \mathbf{u}(x_{p \alpha}) \, \delta(x - x_{p \alpha}) 
	\label{eq005a}
\end{equation}
We devote the following sections to derive its dispersion relationship as function of these general input parameters, enabling the application of the method to a large variety of phononic structures governed Eq. \eqref{eq005a}.
\begin{table}
	\begin{center}
		\begin{tabular}{|l|c|c|}		
    \hline
    Type of scatterer											&  \textbf{Resonators}     & \textbf{Inclusions}  \\
    \hline
	Standing wave equation   &
	\multicolumn{2}{c|}{
	$\displaystyle
	\frac{\textrm{d} \mathbf{u}}{\textrm{d} x} = \mathbf{A} \, \mathbf{u} + 
	\sum_p \sum_{\alpha=1}^{N} \mathbf{K}_\alpha \mathbf{u}(x_\alpha) \, \delta(x - x_{p\alpha}) 	
	$
	}\\ \hline 
	 & & \\
	Matrix $\mathbf{K}_\alpha$	& 
	$	\left[
	\begin{array}{c|c}
	\mathbf{0}  & \mathbf{0} \\
	\hline
	\mathbf{H} \, \left.  \mathbf{T}^{(\alpha)}_{ff}\right. ^{-1} \, \mathbf{T}^{(\alpha)}_{fd} \, \mathbf{H}^T  & \mathbf{0} 
	\end{array} 
	\right]	$ &
	$
	\displaystyle 
	e^{- \mathbf{A} \, \Delta x_\alpha /2 } \ e^{\mathbf{A}_\alpha  \, \Delta x_\alpha /2} - 
	e^{\mathbf{A} \, \Delta x_\alpha /2 } \ e^{-\mathbf{A}_\alpha  \, \Delta x_\alpha /2}$	\\ & & \\
	\hline  
	& & \\
	Sketch & \includegraphics[width=3.5cm]{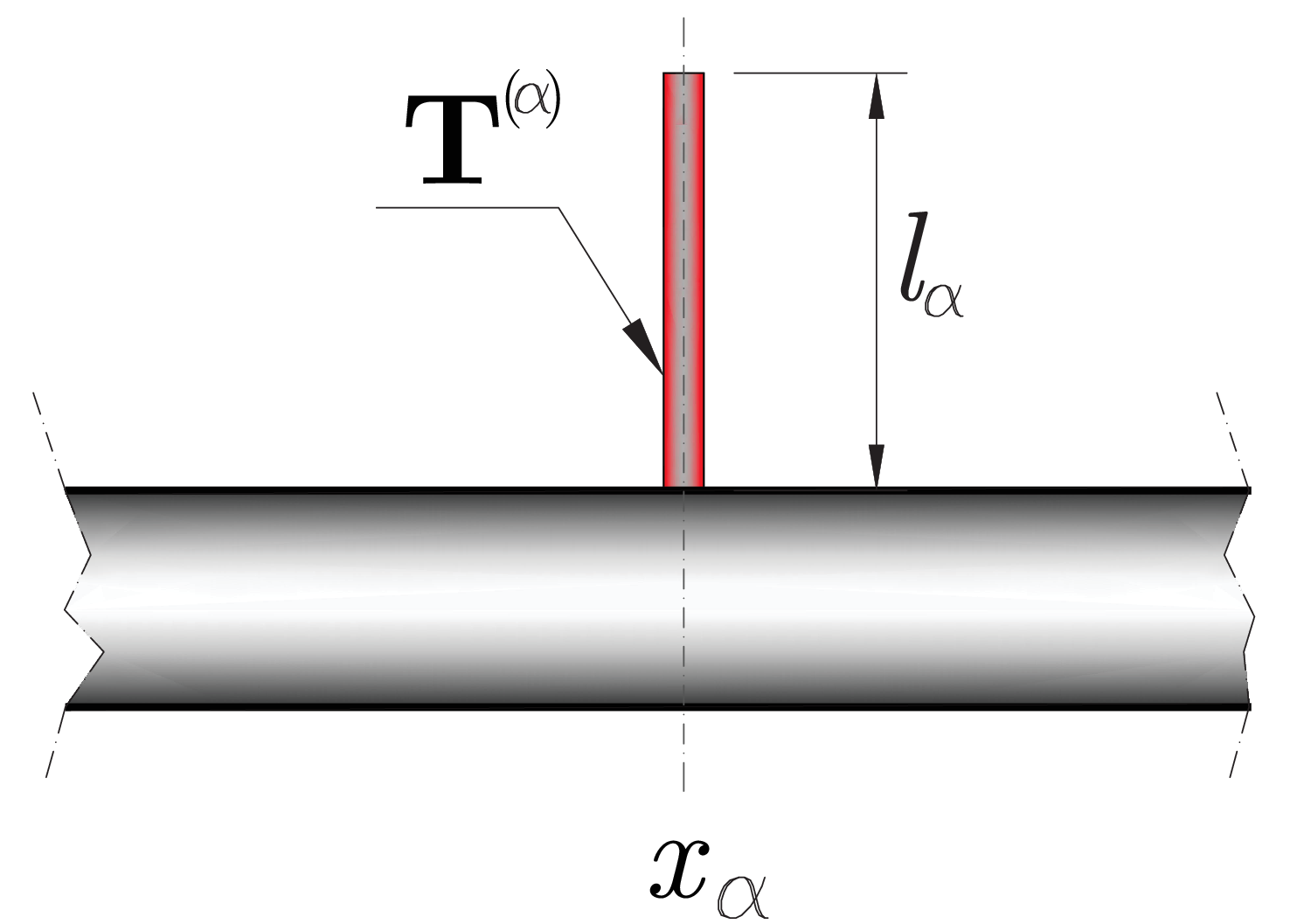} & \includegraphics[width=3.5cm]{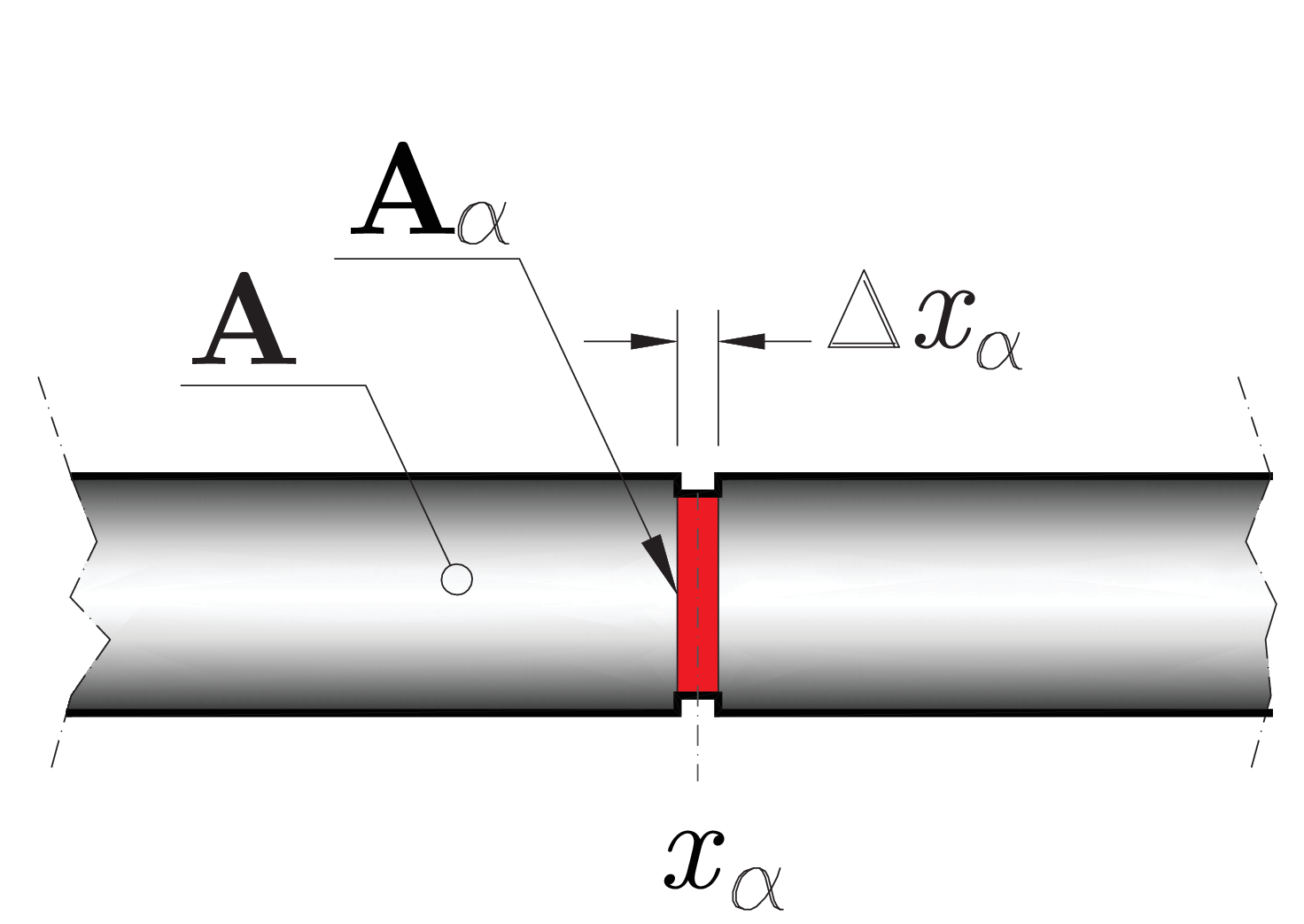} \\ & & \\
	\hline
\end{tabular}
	\end{center}
	\caption{Both resonators and small inclusions in 1D beam-like structures have the same mathematical consideration in the dispersion relations, but under a different form of $\mathbf{K}_\alpha$ matrix. For resonators-type scatterers: matrix $\mathbf{K}_\alpha$ depends on the block-matrices of the transfer matrix $\mathbf{T}^{(\alpha)}$ associated to the whole scatterer and on the matrix $\mathbf{H}$ which relates degrees of freedom of the end of the scatterer with those of the beam-axis (detailed in Sec. 1 from Supplementary Material and in ref. \cite{Lazaro-2024c}). For inclusions-type scatterers: matrix $\mathbf{K}_\alpha$ depends uniquely on both matrices $\mathbf{A}$ and $\mathbf{A}_\alpha$ associated to the properties of the host-medium and scatterer-medium (detailed in Sec. 2 from Supplementary Material  and in ref. \cite{Lazaro-2025a}). }
	\label{tab02}
\end{table}

\section{Dispersion relation}

\subsection{Exact solution}

In this section, we will derive exact equations which define the dispersion relation of any 1D elastic medium loaded by $N$ resonators and governed by the differential equation \eqref{eq005a} in the frequency domain. Our objective is to find the nature of  propagating waves, characterized by a wavenumber $k$ and a waveform $\mathbf{u}(x)$, which in general are related to each other when invoking the Bloch theorem, namely
\begin{equation}
	\mathbf{u}(x) = e^{ikx} \, \bm{\Psi}(x) 
	\label{eq018}
\end{equation}
where $k$ is the wavenumber and  $\bm{\Psi}(x) $ is an unknown periodic function with periodicity $L$ to be found for each value of $k$. Due to the periodicity of the system we can expand in Fourier coefficients yielding 
\begin{equation}
	\bm{\Psi}(x) = \sum_{\sigma = -\infty}^\infty \mathbf{W}_\sigma \, e^{i q_\sigma x} \quad , \quad q_\sigma = \frac{2 \pi \sigma}{L} 
	\ , \quad pL \leq x \leq (p+1)L
		\label{eq019}
\end{equation}
where now the sequence of coefficients $\{\mathbf{W}_\sigma\}_{\sigma=-\infty}^\infty$ are unknown $2m$-size arrays to be found.  Plugging Eqs. \eqref{eq018} and \eqref{eq019} into Eq. \eqref{eq005a}  we obtain after some arrangements
\begin{equation}
	\sum_{\sigma = -\infty}^\infty \left[  i(k + q_\sigma) \mathbf{I} - \mathbf{A}    \right]\mathbf{W}_\sigma  e^{i q_\sigma x} =
	\sum_p \, \sum_{\alpha} \mathbf{K}_\alpha \, \bm{\Psi}(x_{p\alpha}) \, e^{-ik (x - x_{p\alpha})} \, \delta(x - x_{p \alpha}) 
	\label{eq020}
\end{equation}
Considering now an arbitrary integer $\mu$ and $q_\mu = 2 \pi \mu / L$, we can multiply by $e^{- i q_\mu x}$ and integrate within any unit cell, i.e. $\int_{x = pL}^{(p+1)L} \, (\bullet) \, e^{- i q_\mu x} dx$, it yields
\begin{equation}
	\sum_{\sigma = -\infty}^\infty \left[ i(k + q_\sigma) \, \mathbf{I} - \mathbf{A}    \right]\mathbf{W}_\sigma \, \delta_{\mu \sigma} \, L  =
	\sum_{\alpha=1}^N \mathbf{K}_\alpha \, \bm{\Psi}(\xi_{\alpha}) \, e^{- i q_\mu \, \xi_\alpha } \ ,
	\label{eq021}
\end{equation}

where $\delta_{\mu \sigma} $ stands for the Kronecker delta-function, leading to 
\begin{equation}
	\left[  i(k + q_\mu) \mathbf{I} - \mathbf{A}    \right] \mathbf{W}_\mu \, L  =
	\sum_{\alpha=1}^N \mathbf{K}_\alpha \, \bm{\Psi}(\xi_{\alpha}) \, e^{- i q_\mu \, \xi_\alpha } \ .
	\label{eq022}
\end{equation}
Expressing the vectors $\bm{\Psi}(\xi_{\alpha})$ also in terms of the Fourier coefficients as $\bm{\Psi}(\xi_{\alpha}) = \sum_{\sigma = -\infty}^\infty \mathbf{W}_\sigma \, e^{i q_\sigma \xi_{\alpha}} $ we obtain
\begin{equation}
	\left[  i(k + q_\mu) \mathbf{I} - \mathbf{A}    \right] \mathbf{W}_\mu \, L  =
	 \sum_{\sigma = -\infty}^\infty \sum_{\alpha=1}^N \mathbf{K}_\alpha \, \mathbf{W}_\sigma \, e^{ i (q_\sigma - q_\mu) \, \xi_\alpha } \ .
	\label{eq025}
\end{equation}

The above relations represent an eigenvalue problem with eigenvalue $k$ and where the eigenvector is represented by the infinite set $\{\mathbf{W}_\mu\}_{\mu=-\infty}^\infty$.  Since for each index $\mu$, $\mathbf{W}_\mu$ is a vector of $2m$ components, we can decompose it in the spectral basis of matrix $\mathbf{A}$ given by Eqs. \eqref{eq066}, expanding in terms of the right-eigenvectors $\{\mathbf{u}_l\}_{l=1}^{2m}$, resulting in $2m$ coefficients $a_{l}(\mu)$
\begin{equation}
	\mathbf{W}_\mu = \sum_{l=1}^{2m} a_{l}(\mu) \, \mathbf{u}_l \ , 
	\label{eq023}
\end{equation}
or readily more compacted as
\begin{equation}
	\mathbf{W}_\mu = \mathbf{U} \, \mathbf{a}(\mu)  \ , 
	\label{eq024}
\end{equation}
where $\mathbf{U} = \left[\mathbf{u}_{1},\ldots,\mathbf{u}_{2m}\right]$ is the square matrix of size $2m$ with the elements of basis $\{\mathbf{u}_l\}_{l=1}^{2m}$ as columns, and $ \mathbf{a}(\mu) = \{a_1(\mu),\ldots,a_{2m}(\mu)\}^T$ stands for the $2m$ coefficients of $	\mathbf{W}_\mu$ in such basis. Plugging this expression into Eq. \eqref{eq025}
\begin{equation}
	\left[  i(k + q_\mu) \mathbf{I} - \mathbf{A}    \right] \mathbf{U} \, \mathbf{a}(\mu)  \, L  =
	\sum_{\sigma = -\infty}^\infty \sum_{\alpha=1}^N \mathbf{K}_\alpha \, \mathbf{U} \, \mathbf{a}(\sigma)  \, e^{ i (q_\sigma - q_\mu) \, \xi_\alpha }
	\label{eq026}
\end{equation}
Similarly as above we can build the matrix $\mathbf{V} = \left[\mathbf{v}_{1},\ldots,\mathbf{v}_{2m}\right]$ with the left-eigenvectors of $\mathbf{A}$ in columns. 
Multiplying by $\mathbf{V}^T$ and using the orthogonal relations of Eq. \eqref{eq066} in matrix form $\mathbf{V}^T \mathbf{A} \mathbf{U} = i \mathbf{k}_0 = \text{diag}\left[ik_1,\ldots,ik_{2m}\right]$  and $\mathbf{V}^T  \mathbf{U} = \, \mathbf{I}_{2m}$ , we find an equivalent expression to that of Eq. \eqref{eq025} but now with the additional advantage that the lelf-hand side  of the equation is diagonal
\begin{equation}
	iL \left[  (k + q_\mu) \mathbf{I}_{2m} - \mathbf{k}_0    \right]  \mathbf{a}(\mu) = 
		\sum_{\sigma = -\infty}^\infty \sum_{\alpha=1}^N \mathbf{V}^T  \mathbf{K}_\alpha \, \mathbf{U}  \, \mathbf{a}(\sigma) \, e^{ i (q_\sigma - q_\mu) \, \xi_\alpha }
		\ , \qquad	- \infty < \mu <  \infty
	\label{eq027}
\end{equation}
In the above infinite system of equations in the coefficients $\mathbf{a}(\mu), \ - \infty < \mu <  \infty$, we identify an eigenvalue problem in the parameter \(k\), whose eigenvector is formed by sequence of $2m$-sized vectors of coefficients  \(\mathbf{a}(\mu), \  \mu = 0, \pm 1, \pm 2, \ldots \). For its numerical implementation, it is necessary to truncate the previous series by taking a finite number of plane waves. The linear (finite--sized) eigenvalue problem derived from this truncation has been obtained in Sec. 4 from the Supplementary Material. In the context of multiple scattering response of one-dimensional elastic waveguides, the precise definition of what weak scattering means has to do with the eigenvalues of the scattering problem matrix \cite{Lazaro-2024c}, which considers, in the frequency domain, the effect of both the underlying patterns in the scatterers' position and their internal resonances. In the present investigation we seek the precise definition of weak scattering for 1D generalized phononic crystals. Thus, using Eq. \eqref{eq027} as starting point, in the next section a rigorous approach is addressed to this end.

 \subsection{Approximate solutions based on weak scattering}

The introduction of perturbations in the 1D elastic medium modifies the behavior of its dispersion relation. It is well known that in the case of spatially periodic perturbations band gaps emerge, which prevent the wave from propagating in this particular range of frequency. However, the reality is that in the presence of weak scattering, wave propagation is verified at the vast majority of frequencies, even in a slightly different way than in a homogeneous medium. In this section, we are particularly interested in shedding light on how the wave propagation is perturbed in the presence of $N$ scatterers in the unit cell. We do not expect the approximate solutions to accurately reproduce the dispersion relation within bandgaps, since in them precisely the scattering cannot be considered weak due to the absence of wavemode transmission. \\

\begin{figure}[h]%
	\begin{center}
		\begin{tabular}{c}
			\includegraphics[width=13cm]{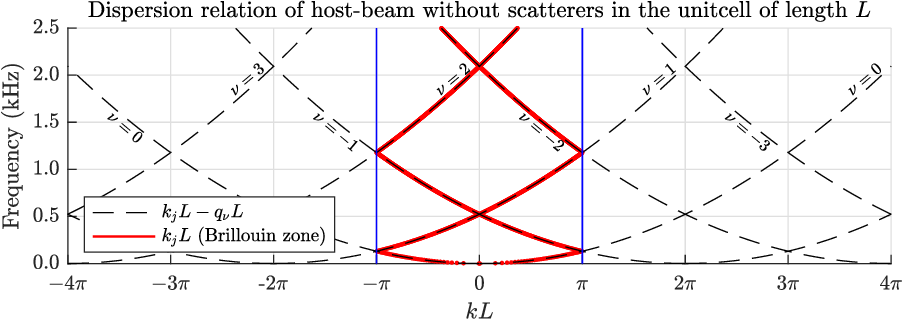} 
		\end{tabular}			
		\caption{Dispersion relations of the 1D elastic homogeneous medium. The different branches correspond to expressions of the form $k_jL - q_\nu$, $q_\nu = 2\pi \nu / L, \ , \ \nu = 0, \pm 1, \pm 2,\ldots$. Within the Brillouin zone, they constitute the same curve as folding the branch $q_0=0$. }%
		\label{fig10}%
	\end{center}
\end{figure}

Looking at general eigenvalue problem of  Eq. \eqref{eq027},  and canceling the effect of scatterers  then the problem reduces to
\begin{equation}
	iL \left[   \mathbf{k}_0  -  q_\mu \mathbf{I}_{2m} \right]  \mathbf{a}_0(\mu) = \lambda \,  \mathbf{a}_0(\mu)
	\quad , \qquad - \infty < \mu < \infty
	\label{eq035}
\end{equation}
where we denote as $\{\mathbf{a}_0(\mu)\}_{\mu=-\infty}^{\infty}$ for the coefficients of the wavemodes corresponding to the bare medium without scatterers.  Composing for all indexes $\mu$, Eq. \eqref{eq035} is a trivial diagonal eigenvalue problem in the parameter $\lambda = ikL$, where the solutions depend on two indexes $(j,\nu)$. Thus, each eigenvalue can be expressed as one element of the infinite set 
\begin{equation}
		\lambda(j,\nu) =  \{ik_jL - i q_{\nu} L : 1 \leq j \leq 2m \ , \ -\infty \leq \nu \leq \infty\}
			\label{eq038}
\end{equation}
with $k_j$ the $j$th wavenumber of the homogeneous medium. And the corresponding eigenvector $\{\mathbf{a}_0(\mu)\}_{\mu=-\infty}^{\infty}$ associated to the eigenvalue $ik_jL - i q_{\nu} L$ is
\begin{equation}
		\mathbf{a}_0(\mu) =  
        \begin{cases}
            \mathbf{0} \in \mathbb{C}^{2m}  & \text{if } \mu \neq \nu \\
            \{0,\ldots,0,\underbrace{1}_{\text{pos. } j},0,\ldots,0\}^T & \text{if } \mu = \nu             
        \end{cases}
		\label{eq037}
\end{equation}
Accounting for these coefficients and following Eqs. \eqref{eq019} and \eqref{eq023}, the eigenmode can then readily be written as
\begin{equation}
	\bm{\Psi}_{0}(\xi) = e^{iq_\nu \xi} \, \mathbf{u}_j
	\label{eq053b}
\end{equation}
Fixed the index $j$, every mode of the form $\lambda = \pm ik_jL - i q_{\nu} L$ for $\nu = 0,\pm 1, \pm 2, \ldots$ represents an additional branch in the  irreducible Brillouin zone corresponding to the bare generalized beam. In order to visualize the meaning of these eigenvalues the different branches  $k_jL - q_\nu L$,  with $-3 \leq \nu \leq 3$ have been plotted in Fig. \ref{fig10} for a steel Euler-Bernouilli beam with a cross section of dimensions 50x100 mm$^2$. The left/right bending propagating modes have respectively the expressions  $k_{1,3} = \pm \left(\varrho A \omega^2 / EI \right)^{1/4}$ (the other two modes $k_{2}$ and $k_4$ are evanescent, see Sec. 3 from Supplementary Material).  Notice that the different branches within the Brillouin zone are equivalent to  folding up that one corresponding to $q_0 = 0$. \\

According to the above and without loss of generality, we fix one of the modes of the unperturbed medium, denoted by $(j,\nu)$. For illustrative purposes, we assume that the arrangement of scatterers induces a slight perturbation in the dispersion relation of Eq.~\eqref{eq038}. 
The components of the perturbed wavemode will be formed by the infinite sequence of arrays  $\{\mathbf{a}(\mu)\}_{\mu=-\infty}^{\infty}$ which are expected to change  slightly with respect to the unperturbed ones of the host-beam, say $\{\mathbf{a}_0(\mu)\}_{\mu=-\infty}^{\infty}$ defined in Eq.~\eqref{eq037}. 	Since we can choose any type of normalization, we will impose the $j$th component of  $\mathbf{a}(\nu) = \{a_1(\nu),\ldots,a_j(\nu),\ldots,a_{2m}(\nu)\}^T$ to be
\begin{equation}
	a_j(\nu) = 1 \quad , \qquad \text{normalization relationship for mode $(j,\nu)$}
	\label{eq039}
\end{equation}
The eigenvalue $\lambda = ikL$ and the eigenvector $\{\mathbf{a}(\mu)\}_{\mu=-\infty}^{\infty}$ are related by means of Eq. \eqref{eq027},  which we rewrite using a more compact form for the right-hand side, yielding
\begin{equation}
	iL \left[  (k + q_\mu) \mathbf{I}_{2m} - \mathbf{k}_0    \right]  \mathbf{a}(\mu) = 
	\sum_{\sigma = -\infty}^\infty \mathbf{H}(\mu,\sigma)  \, \mathbf{a}(\sigma) \quad , \quad  \mu = 0, \pm 1, \pm 2, \ldots
	\label{eq029}
\end{equation}
where the matrices $	\mathbf{H}(\mu,\sigma)$ are defined as
\begin{equation}
	\mathbf{H}(\mu,\sigma) = \sum_{\alpha=1}^N \mathbf{V}^T  \mathbf{K}_\alpha \, \mathbf{U}   \, e^{ i (q_\sigma - q_\mu) \, \xi_\alpha }
	\label{eq028}
\end{equation}
The eigenvalue problem  presented in Eq. \eqref{eq029} cannot be solved analytically in the parameter $\lambda = ikL$ since is formed by an infinite set fully coupled homogeneous. The solution required the truncation of the right-hand side series using  $2M+1$ plane waves, i.e. $-M \leq \sigma \leq M$ and considering only the central $2M+1$ equations corresponding to $-M \leq \mu \leq M$. The procedure is described in detail in Sec. 4 from supplementary material for its numerical implementation, resulting finally in the linear eigenvalue problem 
\begin{equation}
	\left(\hat{\mathbf{H}} + \mathbf{D} \right) \hat{\mathbf{a}} = \lambda \, \hat{\mathbf{a}} 
	\label{eq030}
\end{equation}
where the eigenvector $\hat{\mathbf{a}}$ encapsulates the components $ \{\mathbf{a}(\mu), -M \leq \mu \leq M\}$,  $\mathbf{D}$ is formed by diagonal blocks	$iL  ( \mathbf{k}_0 -  q_{\mu} \mathbf{I}_{2m})$, $-M \leq \mu \leq M$ and $	\hat{\mathbf{H}} $ is a block matrix of size $[2m \cdot (2M+1)]^2$ formed blocks defined in Eq. \eqref{eq028}. However, some analytical expressions can be derived by taking advantage of the remarkable decoupling of the left-hand side terms in Eq. \eqref{eq029}. Ultimately, these expressions will allow us to design an iterative procedure that can, under certain conditions, find the exact solution. Furthermore, the first and second iterations easily lead to expressions that can serve as a suitable analytical solution for problems in the  weak scattering approximation. \\

Let us assume on one hand that, for our fixed mode of indexes $(j,\nu)$, the form of the eigenmode given by the periodic function $ \bm{\Psi}(x)$ is known.  Let us see that the value of $k$ corresponding to the dispersion relation can be explicitly found. Indeed, if we focus on the $j$th relation of Eq. \eqref{eq027} we have
\begin{equation}
	iL \left[  (k + q_\mu  )-  k_j  \right]  a_j(\mu) = 
	 \sum_{\alpha=1}^N  \mathbf{v}_j^T  \mathbf{K}_\alpha \, \bm{\Psi}(\xi_{\alpha}) \, e^{ - i   q_\mu \, \xi_\alpha } \ , \quad \mu = 0, \pm 1, \pm 2, \ldots
	\label{eq040}
\end{equation}
where the mode $\bm{\Psi}(\xi)$ can be expressed, after merging  Eqs. \eqref{eq019} and \eqref{eq024},  as 
 $$\bm{\Psi}(\xi) = \sum_{\sigma = -\infty}^\infty \mathbf{U}  \, \mathbf{a}(\sigma)  \, e^{i q_\sigma \xi}.   \ ,   $$
Using now the normalization relationship $a_j(\nu) = 1$ of Eq. \eqref{eq039}, we can evaluate Eq. \eqref{eq040} at $\mu = \nu$  and readily have the value of $k$ as function of the eigenmode $\bm{\Psi}(\xi)$
\begin{equation}
	k  =  k_j - q_\nu  + \frac{1}{iL}  
	\sum_{\alpha=1}^N  \mathbf{v}_j^T  \mathbf{K}_\alpha \, \bm{\Psi}(\xi_{\alpha}) \, e^{ - i   q_\nu \, \xi_\alpha }
	\label{eq041}
\end{equation}
Now, consider on the other hand that the eigenvalue $k$ is known but not the coefficients of eigenmode, $\mathbf{a}(\mu)$. From Eq. \eqref{eq040} changing the index $j$ by the generic $l$ and solving for $a_l(\mu)$, we find
\begin{equation}
	a_l(\mu) = 
	\sum_{\alpha=1}^N  \mathbf{v}_l^T  \mathbf{K}_\alpha \, \bm{\Psi}(\xi_{\alpha}) \, \frac{e^{ - i   q_\mu \, \xi_\alpha }}{iL \left[  k -  k_l + q_\mu   \right]  }
	\label{eq042}
\end{equation}
The mode is then
\begin{equation}
	 \bm{\Psi}(\xi) = \sum_{\mu=-\infty}^\infty \sum_{l=1}^{2m} a_l(\mu) \mathbf{u}_l e^{iq_\mu \xi} =
	 	\sum_{\mu=-\infty}^\infty \sum_{l=1}^{2m} \mathbf{u}_l 
	 	\left(\sum_{\alpha=1}^N  \mathbf{v}_l^T  \mathbf{K}_\alpha \, \bm{\Psi}(\xi_{\alpha}) \, \frac{e^{ - i   q_\mu \, \xi_\alpha }}{iL \left[  k  -  k_l + q_\mu   \right]  }\right)e^{iq_\mu \xi}  
	 	\label{eq043}					
\end{equation}
After some rearangements, this expression can be compacted as

\begin{equation}
	\bm{\Psi}(\xi) =
	\sum_{\alpha=1}^N 
	{\mathbf{G}}(k, \xi - \xi_\alpha) \mathbf{K}_\alpha \,  \bm{\Psi}(\xi_{\alpha}) 	
	\label{eq050}
\end{equation}
where the function ${\mathbf{G}}\left(k, \xi \right) \in \mathbb{C}^{2m \times 2m}$ is a matrix for each pair $(k,\xi)$ and  is defined as
\begin{equation}
	{\mathbf{G}}(k, \xi) =  \sum_{l=1}^{2m} 
	\mathbf{u}_l \mathbf{v}_l^T  \, \varphi\left(k-k_l, \xi\right) 
	\label{eq045}
\end{equation}
Above, the new function $\varphi(\kappa,\xi)$ captures the waveform periodicity and has the simple expression
\begin{equation}
	\varphi(\kappa,\xi) = \sum_{\mu = -\infty}^\infty \frac{ e^{i q_\mu \xi } } {iL	\left( \kappa +  q_\mu\right)} 
	\label{eq046}
\end{equation}
 which can be simplified further since its sum can be obtained as the periodic expansion by Fourier series to the whole  $x$--axis of the following closed-form expression, defined in the interval $\left[0,L\right]$
\begin{equation}
	\varphi(\kappa,\xi) = 
	\begin{cases}
			\displaystyle \frac{e^{- i\kappa \xi }}{1 - e^{-i \kappa L }} &  0 \leq \xi \leq L  \\
			\varphi(\kappa,\xi - p L) & pL \leq \xi \leq (p+1)L \ , \ p = \pm 1, \pm 2, \ldots
	\end{cases} 
	\label{eq048}
\end{equation}
The notation of the function $\mathbf{G}(k,x)$ has not been chosen arbitrarily since it is closely related to the Green function of the 1D homogeneous medium under certain configuration of applied forces. Indeed, consider that at the locations $x_p = pL, \ , \quad p = 0,\pm 1, \pm 2, \ldots$ a source of magnitude $e^{ikx_p}$ is applied. The Green function in such situation is the $k$-dependent matrix $\bm{\mathcal{U}}(k,x)$ which verifies the following differential equation
\begin{equation}
	\left( \frac{\textrm{d}}{\textrm{d} x} -  \mathbf{A} \right) \bm{\mathcal{U}} = \sum_{p=-\infty}^\infty \mathbf{I} \, e^{ikpL}\delta (x - pL)
	\label{eq070}
\end{equation}
Then, it can be proved that this Green function is indeed $\bm{\mathcal{U}}(k,x) = e^{ikx}\mathbf{G}(k,x)$, where $\mathbf{G}(k,x) $ has been defined in Eq. \eqref{eq045}. The matrix $ \mathbf{G}(k,x)$ stands for the periodic response of the homogeneous medium to a periodic distribution of point sources every $L$ meters.\\

Evaluating now Eq. \eqref{eq050} at $\xi = \xi_1,\ldots,\xi_N$ we find the following system of $2mN$ algebraic equations
\begin{equation}
		\left\{
	\begin{array}{c}
	\bm{\Psi}(\xi_{1}) 	  \\  \bm{\Psi}(\xi_{2}) 				 \\ \vdots \\ 			\bm{\Psi}(\xi_{N}) 	
	\end{array}\right\}	=
	\left[
	\begin{array}{cccc}
		\mathbf{G}(k,0)\mathbf{K}_1			&			\mathbf{G}(k,\xi_1-\xi_2) \mathbf{K}_2   & \cdots 	&  \mathbf{G}(k,\xi_1-\xi_N) \mathbf{K}_N \\
		\mathbf{G}(k,\xi_2-\xi_1) \mathbf{K}_1  &  		\mathbf{G}(k,0)\mathbf{K}_2	   & \cdots 	&  \mathbf{G}(k,\xi_2-\xi_N) \mathbf{K}_N \\
		\vdots					 & 				\vdots															 &   \ddots  &   \vdots  \\
		\mathbf{G}(k,\xi_N-\xi_1) \mathbf{K}_1			&	\mathbf{G}(k,\xi_N-\xi_2) \mathbf{K}_2   & \cdots 	& 			\mathbf{G}(k,0)\mathbf{K}_N
	\end{array}
	\right]
	\left\{
\begin{array}{c}
	\bm{\Psi}(\xi_{1}) 	  \\  \bm{\Psi}(\xi_{2}) 				 \\ \vdots \\ 			\bm{\Psi}(\xi_{N}) 	
\end{array}\right\}	
	\label{eq051}
\end{equation} 
Introducing the new notation
\begin{equation}
		\hat{\bm{\Psi}} =
	\left\{
	\begin{array}{c}
		\bm{\Psi}(\xi_{1}) 	  \\  \bm{\Psi}(\xi_{2}) 				 \\ \vdots \\ 			\bm{\Psi}(\xi_{N}) 	
	\end{array}\right\} , \ 
	\hat{\mathbf{G}}(k) =
	\left[
	\begin{array}{cccc}
		\mathbf{G}(k,0)	&			\mathbf{G}(k,\xi_1-\xi_2)  & \cdots 	&  \mathbf{G}(k,\xi_1-\xi_N) \\
		\mathbf{G}(k,\xi_2-\xi_1)  &  		\mathbf{G}(k,0)   & \cdots 	&  \mathbf{G}(k,\xi_2-\xi_N)  \\
		\vdots					 & 				\vdots															 &   \ddots  &   \vdots  \\
		\mathbf{G}(k,\xi_N-\xi_1) 	&	\mathbf{G}(k,\xi_N-\xi_2)    & \cdots 	& 			\mathbf{G}(k,0) 
	\end{array}
	\right]
	 , \ 
	\hat{\mathbf{K}} = 
	\left[
	\begin{array}{cccc}
	\mathbf{K}_1	&	\cdots 	&  \mathbf{0}_{2m}  \\
	\vdots					 &   \ddots  &   \vdots  \\
	\mathbf{0}_{2m}  & \cdots 	& 		\mathbf{K}_N
	\end{array}
	\right]
	\label{eq054}
\end{equation} 
we find that the vector $\hat{\bm{\Psi}} $ is obtained as function of the eigenvalue $k$, solving the null subspace associated to the unitary eigenvalue of matrix $\hat{\mathbf{G}}(k) \hat{\mathbf{K}}$, i.e.
\begin{equation}
	\left[\mathbf{I}_{2mN} - \hat{\mathbf{G}}(k) \, \hat{\mathbf{K}} \right] \hat{\bm{\Psi}} = 	\mathbf{0}
	\label{eq052}
\end{equation}
%
As shown above, we can express the wavenumber $k$ as function of $\bm{\Psi}(x)$ in Eq. \eqref{eq041} and inversely the wavemode $\bm{\Psi}(x)$ as function of $k$ from Eq. \eqref{eq050} just solving the null-space problem \eqref{eq052}. And more important is the fact that these expressions are exact and expressed in terms of the eigenmode evaluated at the scatterers. In what follows we will exploit these new relations to generate an iterative scheme. Fixed a mode of the unperturbed beam, say $(j,\nu)$, consider the following initial values, corresponding to the bare waveguide 
\begin{equation}
	k^{(0)} = k_j - q_\nu \qquad , \qquad  \bm{\Psi}_{0}(\xi) = e^{iq_\nu \xi} \, \mathbf{u}_j
	\label{eq053}
\end{equation}
The values of $k^{(0)}$ have been plotted in Fig. \ref{fig10} for an Euler--Bernouilli beam, showing the different branches which arise after evaluating $q_\nu = 2\pi \nu / L$, for $\nu = 0, \pm 1, \pm 2, \ldots$. We define the following sequences $\{k^{(n)}\}$ and $\{\bm{\Psi}_{n}(\xi)\}$ for $n=1,2,\ldots$ 
\begin{eqnarray}
	k^{(n)}  &=&  k_j - q_\nu  + \frac{1}{iL}  
	\sum_{\alpha=1}^N  \mathbf{v}_j^T  \mathbf{K}_\alpha \, \bm{\Psi}_{n-1}(\xi_{\alpha}) \, e^{ - i   q_\nu \, \xi_\alpha } \label{eq054a} \\
	\bm{\Psi}_n(\xi) &=& \sum_{\alpha=1}^N 
	{\mathbf{G}}  \left(	k^{(n)}, \xi - \xi_\alpha\right) \mathbf{K}_\alpha \,  \bm{\Psi}_{n-1}(\xi_{\alpha}) 	
	\label{eq054b}	
\end{eqnarray}
According to the presented scheme, each mode $(j,\nu)$, with $1 \leq j \leq 2m$ and $-M \leq \nu \leq M$ can be obtained by perturbation, following the recursive method presented above. However, as shown in Fig. \ref{fig10}, the modes for $\nu = \pm 1, \pm 2, \ldots$ are just branches that can be obtained folding that of $\nu=0$ in the irreducible Brillouin zone. Therefore we can just keep working with the mode $(j,0)$ or shortly the $j$th mode. Moreover, since only the evaluation of the wavemodes at the scatterers, say $\bm{\Psi}_{n-1}(\xi_{\alpha})$, is needed we can rewrite the above sequences in a more compact form using the structure of the extended vector $ \hat{\bm{\Psi}}$ and matrices $ \hat{\mathbf{G}}(k)$ and $\hat{\mathbf{K}} $ defined in Eq. \eqref{eq054}, resulting
\begin{eqnarray}
	k^{(n)}  &=&  k_j + \frac{1}{iL}   \left(\mathbf{1}^T_N \otimes \mathbf{v}_j^T \right)  \hat{\mathbf{K}} \, \hat{\bm{\Psi}}_{n-1} \label{eq062a} \\
	 \hat{\bm{\Psi}}_{n} 	 &=& \hat{\mathbf{G}} \left(k^{(n)} \right) \, \hat{\mathbf{K} }\, 	 \hat{\bm{\Psi}}_{n-1}
	\label{eq062b}	
\end{eqnarray}
where we use the notation $\mathbf{1}^T_N \otimes \mathbf{v}_j^T $ to represent the row-vector formed by $\mathbf{v}_j^T$, replicated $N$ times. The vector $\mathbf{1}_N = \{1,\ldots,1\}^T$ of size $N$ denotes the column vector of size $N$ formed only by ones. \\

First and second order iterations have a qualitative value since analytical expressions can be derived. Indeed the first order perturbation of the mode $(j,0)$ leads to
\begin{equation}
       	k^{(1)}  =  k_j  + \frac{1}{iL}  
       \sum_{\alpha=1}^N  \mathbf{v}_j^T  \mathbf{K}_\alpha \,  \mathbf{u}_j
       \label{eq071}
\end{equation}
The second iteration is now proportional to the first-order mode, yielding
\begin{equation}
		k^{(2)}  =  k_j  + \frac{1}{iL}  
	\sum_{\alpha=1}^N  \mathbf{v}_j^T  \mathbf{K}_\alpha \, \bm{\Psi}_{1}(\xi_{\alpha}) \label{eq072} 
\end{equation}
where the vectors $		\bm{\Psi}_1(\xi_\alpha)$ can be evaluated from the expression of the first iteration
\begin{eqnarray}
		\bm{\Psi}_1(\xi_\alpha) &=& \sum_{\beta=1}^N 
	{\mathbf{G}}  \left(	k^{(1)}, \xi _\alpha- \xi_\beta \right) \mathbf{K}_\beta \,  \bm{\Psi}_{0}(\xi_{\beta}) 	 \nonumber \\
  &=&	\sum_{\beta=1}^N  
	\sum_{l=1}^{2m} 
	\mathbf{u}_l \mathbf{v}_l^T  \, \varphi\left(k^{(1)} -k_l,  \xi_\alpha  - \xi_\beta \right) 
	\mathbf{K}_\beta \,   \mathbf{u}_j 
	\label{eq072b}
\end{eqnarray}
Plugging this expression into Eq. \eqref{eq072} and considering the particular case with identical scatterers, say $\mathbf{K}_\alpha = \mathbf{K}_0, \ 1 \leq \alpha \leq N$, then, after some rearrangements we have
\begin{equation}
	k^{(2)}  =  k_j  + \frac{1}{iL}  
	\sum_{l=1}^{2m}   (\mathbf{v}_j^T  \mathbf{K}_0 \mathbf{u}_l) (\mathbf{v}_l^T \mathbf{K}_0 	\mathbf{u}_j) \, \zeta \left(k^{(1)}-k_l\right) 
	\label{eq073}
\end{equation}
where the function $\zeta(k)$ is defined as 
\begin{equation}
	\zeta(k) = \sum_\alpha \sum_\beta  \varphi\left( k,  \xi_\alpha  - \xi_\beta \right) 
	\label{eq074}
\end{equation}
Using the periodic expansion of $\varphi\left( k,  \xi \right)  $ given in Eq. \eqref{eq046} this function can be expressed in series as 
\begin{equation}
			\zeta(k) = \sum_{\nu = -\infty}^\infty \frac{ N S(q_\nu) }  {iL	\left(k +  q_\nu\right)} 
	\label{eq075}
\end{equation}
where $S(q)$ stands for the structure factor of the distribution of points \cite{Ashcroft-1976} 
\begin{equation}
	S(q) = \frac{1}{N} \sum_{\alpha=1}^{N} \sum_{\beta=1}^{N}  \, \, e^{- i q \left(\xi_\alpha - \xi_\beta\right)}
	\label{eq076}
\end{equation}
The expression of the 2nd order solution, i.e. Eq. \eqref{eq073}, provides significantly more information than Eq. \eqref{eq071}. It considers the impedance of the scatterers and their impact on the dispersion relation of the mode $j$ by accounting for all modes present in the beam through the coupling coefficients $\mathbf{v}_j^T  \mathbf{K}_0 \mathbf{u}_l$ and $\mathbf{v}_l^T \mathbf{K}_0 	\mathbf{u}_j$. Additionally, the weight factor given by the function $\zeta(k)$ links the spatial correlation of scatterers positions via the Structure Factor, to the scattering properties of the medium. The $\zeta(k)$ function is singular at the coordinates of the reciprocal space, allowing us to interpret the factor $\zeta(k^{(1)} - k_l)$ as an interference coefficient in the mode  $j $ when an incident wavenumber $k_l$  propagates through the medium. If the difference between the scattered wavenumber $k^{(1)}$ and the incident wavenumber $k_l$ corresponds to a crystal lattice node in the reciprocal space, the interference is maximized. This represents the von Laue and the Ewald sphere formulations for 1D phononic crystals.

\section{Convergence of the iterative scheme}
\label{convergence}

Before validating the expressions obtained in the previous section, it is worth asking under what conditions the convergence of the iterative scheme occurs. This is relevant in order to predict the behavior over a range of frequencies. Recalling the form of the sequences, it is straightforward that the sequence of wavenumbers $\{k^{(n) } \}$ will converge provided that the sequence of wavemodes does. In order to derive some mathematical condition for the convergence of this procedure, it turns out that in the Eq. \eqref{eq062b}, the recursive sequence of eigenmodes can be reduced to a fixed-point  iterative scheme. Indeed, consider the following function $\phi: \mathbb{C}^{2mN} \to \mathbb{C}  $ associated to the unperturbed mode $(j,0)$
\begin{equation}
	\phi(\mathbf{X}) = k_j   + \frac{1}{iL}  \left(\mathbf{1}^T_N \otimes \mathbf{v}_j^T \right)  \hat{\mathbf{K}}  \, \mathbf{X}  
	\quad , \quad \mathbf{X} \in  \mathbb{C}^{2mN} 
\end{equation}
It is then clear that for each iteration we obtain $k^{(n)} = 	\phi(\hat{\bm{\Psi}}_{n-1})$.  Let us define the vector self--mapping $\bm{F}(\mathbf{X}):  \mathbb{C}^{2mN} \to  \mathbb{C}^{2mN}$ as
\begin{equation}
		\bm{F}(\mathbf{X}) =   \hat{\mathbf{G}} \left[	\phi(\mathbf{X})  \right] \, \hat{\mathbf{K}} \, \mathbf{X}	
		\label{eq055}	
\end{equation}
 which transforms each (column) array of size $\mathbf{X} \in   \mathbb{C}^{2mN} $ into the same space. Starting with the unperturbed mode $\hat{\bm{\Psi}}_{0} = \mathbf{1}_N \otimes \mathbf{u}_j$, the sequence of eigenmodes $ \{\hat{\bm{\Psi}}_{n}\}_{n\geq 1}$ can be generated just evaluating
 \begin{equation}
 	\hat{\bm{\Psi}}_{n} = \bm{F} \left(\hat{\bm{\Psi}}_{n-1}\right)  \quad , \qquad \hat{\bm{\Psi}}_{0} = \mathbf{1}_N \otimes \mathbf{u}_j
 	\label{eq063}	
 \end{equation}
 This scheme is locally convergent to a fixed--point, say $ 	\hat{\bm{\Psi}}$, if the spectral radius of the Jacobian  matrix of the transformation $\bm{F}(\mathbf{X})$, evaluated at the fixed-point $\mathbf{X} = 	\hat{\bm{\Psi}}$,  is less than the unity \cite{Kirk-2001}, i.e.
 \begin{equation}
 	\rho \left( \mathbf{J} \right) < 1 \quad , \qquad \text{at } \mathbf{X} = 	\hat{\bm{\Psi}}
 	\label{eq064}
 \end{equation}
 where
 and
 	 \begin{equation}
 		\mathbf{J} =  \frac{\partial \mathbf{F} }{\partial \mathbf{X}}
 		\label{eq064c}
 	\end{equation}
 	denotes the Jacobian matrix  of the the self-mapping transformation defined by Eq. \eqref{eq055}, evaluated as the matrix whose elements are the derivatives of its components with respect to the vector $2mN$--sized vector $\mathbf{X}$. And  $\rho(\cdot)$ denotes the spectral radius of a matrix, which as known \cite{Householder-1964}, is the maximum eigenvalue in absolute value , i.e. 
 	\begin{equation}
 		\rho(\mathbf{J}) = \max \{\left|  \lambda \right|: \textnormal{$\lambda$ is eigenvalue of $\mathbf{J}$} \}
 		\label{eq064b}
 	\end{equation}
 	It turns out that  the expression of the Jacobian matrix is available analytically just using the chain rule in the differentiation and being consistent with the dimensions of the resulting expression to obtain a valid matrix, resulting in the matrix
 \begin{equation}
		\mathbf{J} = \frac{\partial \mathbf{F} }{\partial \mathbf{X}} 
		=\frac{1}{iL}  \left[  \frac{\partial \mathbf{G}\left[ 	\phi(\mathbf{X})  \right] }{\partial k} \cdot \hat{\mathbf{K}} \right]\, \cdot 
		\left[\mathbf{X}    \left(\mathbf{1}^T_N \otimes \mathbf{v}_j^T \right) \right] \cdot \hat{\mathbf{K}}  + 
		 \hat{\mathbf{G}} \left[ 	\phi(\mathbf{X})  \right] \, \hat{\mathbf{K}} 
		 \label{eq065}
 \end{equation}
 This matrix results to be a good indicator of the validity of weak--scattering based approaches.  In the numerical examples, its value can be visualized for the frequency ranges studied. In general, the scheme converges quite quickly at the points where Eq. \eqref{eq064} holds. Even at points where it diverges locally, the first two iterations presented in the previous section usually give satisfactory results. \\

 To validate the proposed analytical framework, we present three numerical examples involving different types of phononic crystals: (i) an Euler-Bernoulli beam with discrete spring-mass resonators, (ii) a Timoshenko beam coupled with continuous resonators and (iii) a waveguide with small inclusions. These configurations have been chosen to illustrate the method's versatility across discrete, continuous, and geometric scattering mechanisms. Due to length constraints, the third example has been moved to the Supplementary Materials. This test case, involving waveguides with localized variations in material and geometric properties, is particularly relevant to assess the method's performance in the presence of small inclusions. The results confirm that the analytical expressions remain accurate even under moderate impedance contrast, capturing both propagating and evanescent modes outside the bandgap regions. Moreover, the spectral radius of the Jacobian matrix ---used as a convergence indicator--- remains below unity across most of the studied frequency range, ensuring the reliability of the iterative scheme.

\section{Example 1: 1D phononic crystal with with point-resonators. Flexural waves}

\begin{figure}[h]%
	\begin{center}
					\begin{tabular}{cc}
						\includegraphics[width=9cm]{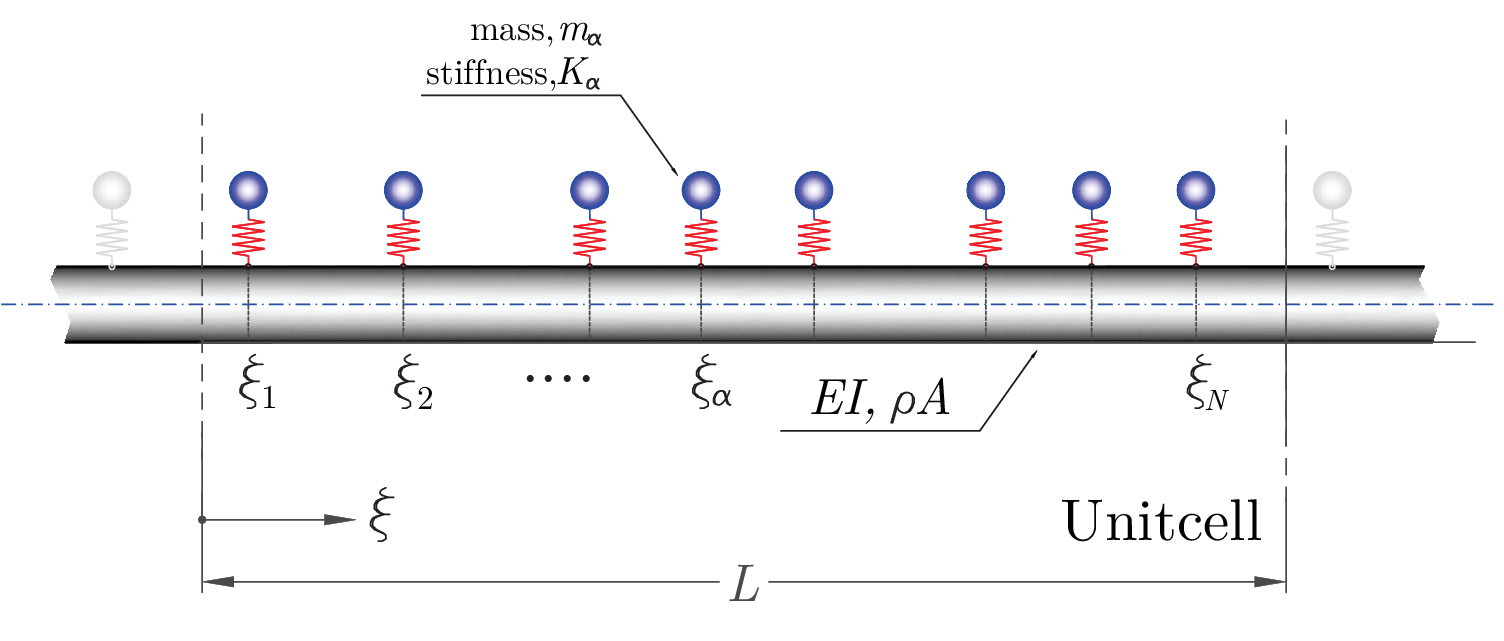} \\						
					\end{tabular}			
		\caption{Unit cell for the 1D phononic crystal considered in Example 1. An Euler-Bernouilli beam with point-resonators formed by one resonance (simple spring-mass system). }%
		\label{fig02}%
	\end{center}
\end{figure}

In this first example, we will study the validity of the proposed expressions for flexural waves traveling in a phononic crystal formed by an Euler-Bernoulli beam, with matrix $\mathbf{A}$ shown in Table \ref{tab01}. We consider $N=5$ simple resonators formed by a mass-spring system and located at the coordinates
$$
\xi_\alpha/L = \{0.224, \    0.471, \     0.525, \     0.774,   \  0.953\}
$$
within a unit cell of length $L=1$ m. The dynamical properties of the beam are a mass per unit of length of $\varrho A = 21$ kg/m, a sectional stiffness of $EI_y = 583$ m$^2$kN. Each resonator has mass of $m_\alpha = 0.3$ kg, spring coefficient $K_\alpha = 350$ kN/m,  and resonance of $\omega_\alpha = \frac{1}{2\pi} \sqrt{K_\alpha / m_\alpha} = 5.4$ kHz. It is known that among the four wavemodes obtained from the dispersion relation of the bare beam, two are propagating and two evanescent. The matrix $\mathbf{A}$ has now size $4 \times 4$ and its corresponding eigenvalues and eigenvectors, say $\lambda_j = i k_j$, $\mathbf{u}_j$ and $\mathbf{v}_j$
are sorted so that the two first represents the two traveling rightwards  and the two last modes are evanescent, resulting in the wavenumbers
\begin{equation}
	k_1 = - \kappa_f 	\ , \quad
	k_2 = + i \kappa_f 	\ , \quad	
	k_3 = + \kappa_f 	\ , \quad		
	k_4 = - i \kappa_f 	\ , \quad    
    \label{eq067}
\end{equation}
 where $\kappa_f = \left(\omega^ 2 \varrho A / EI \right)^{1/4}$. Modes $k_1$ and $k_3$ have propagation nature, while $k_2$ and $k_4$ are pure imaginary (evanescent). Eigenvectors are listed in the Supplementary Material. Each resonator exerts a single vertical force on the beam proportional to the vertical displacement. We find only one nonzero element in the matrix $\mathbf{K}_\alpha$ , resulting the expression
\begin{equation}
	\mathbf{K}_\alpha = m_\alpha  \frac{  \omega_\alpha^2  \, \omega^2}{ \omega^2 -  \omega_\alpha^2}
	\begin{bmatrix}
		0 & 0 & 0 & 0 \\
		0 & 0 & 0 & 0 \\
		 1 & 0 & 0 & 0\\
		0 & 0 & 0 & 0 						
	\end{bmatrix}  
    \label{eq068}
\end{equation}
\begin{figure}[h]%
	\begin{center}
		\begin{tabular}{ccc}
			\multicolumn{3}{c}{\includegraphics[width=9cm]{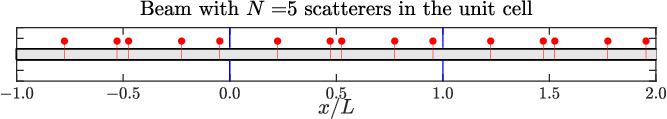}  		} \\ \\
			\includegraphics[width=5cm]{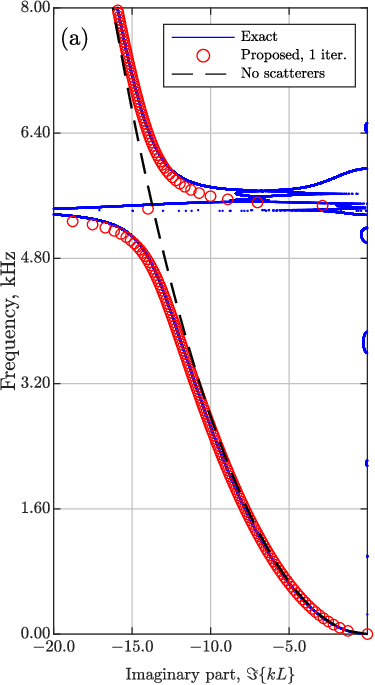} &
			\includegraphics[width=5cm]{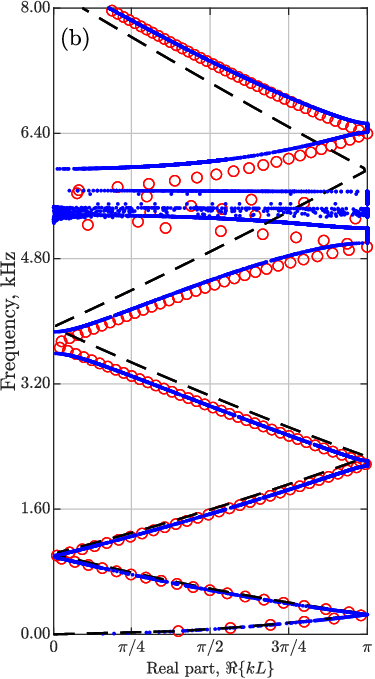} &
			\includegraphics[width=5cm]{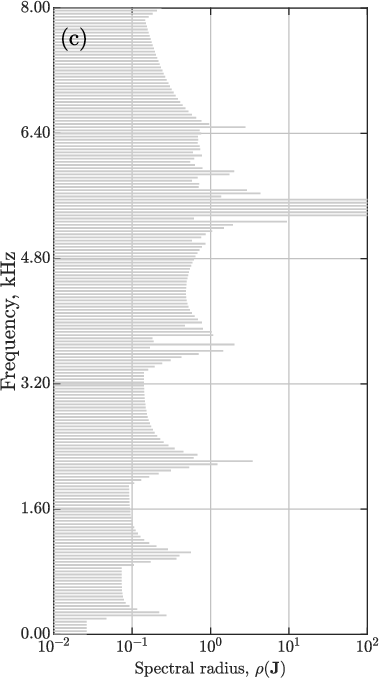} \\			
		\end{tabular}			
		\caption{1D phononic crystal formed by a beam (flexural waves) with $N=5$ discrete resonators at positions $\xi_\alpha/L = \{0.224, \    0.471, \     0.525, \     0.774,   \  0.953\}$. (a) and (b) Imaginary and real part of dispersion relation in Example 1, respectively. Exact result (blue), Iterative approach with the first-order iteration (red markers) and unperturbed homogeneous solution (dashed-black). (c) Plot of the spectral radius of the Jacobian matrix $\mathbf{J}$, criterium of convergence is $\rho(\mathbf{J})<1$}%
		\label{fig04}%
	\end{center}
\end{figure}

As seen in the theoretical derivations, the scatterers' information (position and dynamic properties) and the modes of the bare beam are sufficient to estimate in closed form the new dispersion relation as a perturbation of the four modes shown in Eq. \eqref{eq067}.
In what follows we will obtain the closed forms resulting from the first and second order perturbations of an arbitrary mode $j$ with unperturbed wavemode $k_j$. For the rest of  modes, analogue expressions can be found. Thus, the first order iteration can be found explicitly from Eq. \eqref{eq071}, and using analytical expressions of eigenvectors given in Sec. 3 of the Supplementary Material,  yielding for this particular case
	\begin{equation}
			k^{(1)} =  k_j  \left(1 -  \sum_{\alpha} \frac{m_\alpha}{4\varrho A L} \ \frac{  \omega_\alpha^2}{ \omega^2 -  \omega_\alpha^2}  \right)
	\label{eq069}
\end{equation}
which in the case of identical scatterers with $m_\alpha = m_r$ and $\omega_\alpha = \omega_r$, yields
		\begin{equation}
					k^{(1)} = 	k_j  \left(1 -  \frac{N m_r}{4\varrho A L} \ \frac{  \omega_r^2}{ \omega^2 -  \omega_r^2}  \right)
			\label{eq077}
		\end{equation}
where the wavennumber of the unperturbed mode $k_j$ is given in Eq. \eqref{eq067}. The results of this expression evaluated for the two modes 1 and 2, with $k_1 = -\kappa_{f}$ and $k_2 = i\kappa_f$, have been plotted  in Fig. \ref{fig04}(a) and Fig. \ref{fig04}(b), respectively. The latter is obtained by folding Eq. \eqref{eq077} into the irreducible Brillouin zone. The derived expression is continuous for all frequencies except at the scatterer resonances. It accounts for the number of scatterers, $N$, but not their distribution. Generally, the expression captures the change in wave velocity in regions away from the resonances, respect to the behaviour without resonators. It can also be seen how the model is also able to predict the dispersion of the evanescent mode, see Fig. \ref{fig04}(a) and how it is affected by the resonance, resulting in a fairly accurate prediction. The exact result of the dispersion relation has been obtained using the transfer matrix method for elastic phononic crystals, evaluated along the unit cell \cite{Lazaro-2022a,Davies-2024,Rui-2019}. \\

Figure \ref{fig04}(c) illustrates the spectral radius of the Jacobian matrix, \(\rho(\mathbf{J})\), as derived in Eq. \eqref{eq065}. This spectral radius is crucial for assessing the convergence of the iterative method. When \(\rho(\mathbf{J})\) is less than 1, the method converges locally to the exact mode, indicating that it is attractive. However, if \(\rho(\mathbf{J})\) exceeds 1, the exact mode becomes a repulsive fixed point, preventing the method from reaching the accurate value. This parameter reflects the influence of both Bragg peaks and model resonances. Regions where \(\rho(\mathbf{J})\) is greater than 1 are close to bandgaps and the results of the approximate expressions become less accurate since recursive method diverge after a few iterations. \\

\begin{figure}[h]%
	\begin{center}
		\begin{tabular}{ccc}
			\multicolumn{3}{c}{\includegraphics[width=9cm]{figures/figure_Example01_Sketch.eps}  		} \\ \\
			\includegraphics[width=5cm]{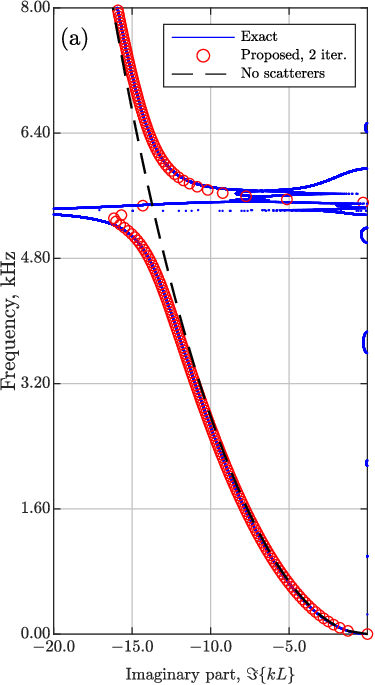} &
			\includegraphics[width=5cm]{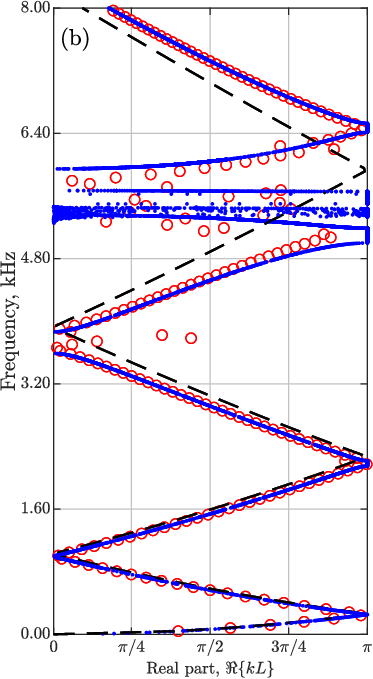} &
			\includegraphics[width=5cm]{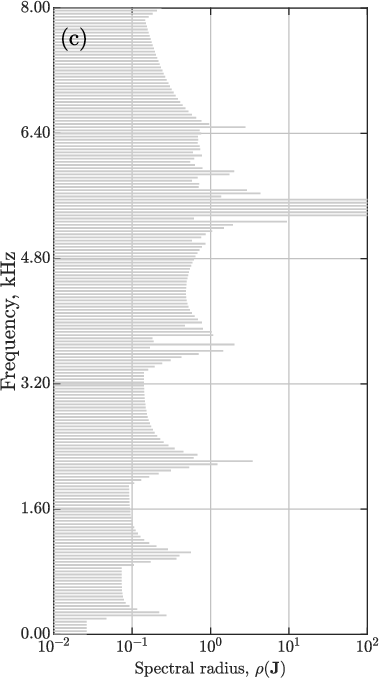} \\			
		\end{tabular}			
		\caption{(a) and (b) Imaginary and real part of dispersion relation in Example 1, respectively. Exact result (blue), Iterative approach with the second-order iteration (red markers) and unperturbed homogeneous solution (dashed-black). (c) Plot of the spectral radius of the Jacobian matrix $\mathbf{J}$, criterium of convergence is $\rho(\mathbf{J})<1$}%
		\label{fig05}%
	\end{center}
\end{figure}

The effect of the scatterer distribution and the interaction of the perturbed mode with the other modes is captured by the second-order approximation derived in Eqs. \eqref{eq072} and \eqref{eq073}. Applying this expression to the current example, where all scatterers are identical, we obtain after some rearrangements the following expression associated to the first propagating mode $k_1 = -\kappa_f$
\begin{equation}
	k^{(2)} =  - \kappa_{f} + \frac{1}{L}\left(\frac{\kappa_{f} m_r}{4\varrho A } \ \frac{  \omega_r^2}{ \omega^2 -  \omega_r^2} \right)^2
		\left[i \zeta \left(k^{(1)} + \kappa_f \right) - i \zeta \left(k^{(1)} - \kappa_f \right)  +  \zeta \left(k^{(1)} - i \kappa_f \right) -  \zeta \left(k^{(1)} + i \kappa_f \right)\right]
	\label{eq078}
\end{equation}
while that corresponding to the evanescent mode $k_2 = i \kappa_f$ is just directly multiplying the previous expression by $-i$, i.e.  given by $-ik^{(2)}$. Above, \(\zeta(k)\) represents the function introduced in Eq. \eqref{eq075}, which relates the distribution of points to the wavenumber of the medium through the structure factor. This function is singular at the coordinates of the medium's reciprocal space. In the specific case of Eq.\eqref{eq078}, this property implies that at frequencies where the difference between the scattered wavenumber and that of the homogeneous medium matches a node of the reciprocal crystal lattice, scattering and interference reach their maximum (von Laue formulation). This property is shown in Fig. \ref{fig05}(b), where the effect of these singularities emerges close to the frequencies associated to the Bragg peaks of the medium. It is observed that the second-order approximation provides a better fit to the exact dispersion relation in regions where \( \rho(\mathbf{J}) < 1\). 	In the region around the resonance at 5.4 kHz, values of the spectral radius greater than one are detected, and this leads to divergence in the proposed iterative sequence. For this reason, comparing Figs. \ref{fig04}(a) and \ref{fig05}(a), the first-order approximation of the evanescent mode is apparently more accurate than that of the second-order. Precisely the latter shows signs of divergence in this region with \( \rho(\mathbf{J}) > 1\). \\

\section{Example 2:  1D phononic crystals with attached beams. Full wave problem.}

In this example, we examine a waveguide incorporating \( N = 10 \) beam-resonators, designed to vibrate both transversely and longitudinally. The host beam is made of aluminum with a square cross-section of dimensions \( h \times b = 12 \times 12 \, \text{cm} \) cm cross-section Timoshenko beam  that reproduces both flexural and shear waves. The properties of the bare beam without perturbations are
\begin{equation}
	EA = 10.08 \times 10^5 \ \text{kN}  , \ 
	EI =  1210 \ \text{kNm$^2$} , \
	GA = 2.45 \times 10^5 \ \text{kNm}  , \
	\varrho A = 30.2 \ \text{kg/m}  , 
	\varrho I = 0.036 \ \text{kg m$^2$/m}, \
	\label{eq080}	
\end{equation}
with this parameters the cutoff frequency of the beam is $\omega_c = \frac{1}{2\pi}\sqrt{\frac{GA}{\rho I_y}} = 13$ kHz, which separates the frequency range into two regions. For frequencies $\omega < \omega_c$, the medium exhibits a propagating mode associated with flexural waves, which, for very low frequencies, say $\omega < 0.05 \omega_c$, coincides with the dispersion of the Euler-Bernoulli beam. Conversely, for $\omega > \omega_c$, in addition to the flexural mode, the shear deformation mode appears. The resonators, also aluminum, feature a cross-section of \( h_\alpha \times b_\alpha = 2.4 \times 2.4 \, \text{cm} \) and a length of \( l_\alpha = 24 \, \text{cm} \). Table \ref{tab04} provides the natural frequencies of the resonators' longitudinal and flexural vibration modes.

\begin{table}[h]	
	\begin{center}
		\begin{tabular}{lcccc}
										 &   $\omega_1 / \omega_c$  &  $\omega_2 / \omega_c$	&  $\omega_3 / \omega_c$      \\ \hline 
Longitudinal modes   &	0.460									    &  1.380 										&  2.300  \\
Flexural modes   &	0.030									    &  0.186 										&  0.521  \\
\hline
\end{tabular}
		\caption{Natural frequencies of the beam-resonators relative to the cutoff frequency of the host-beam}
		\label{tab04}
	\end{center}	
\end{table}

\begin{figure}[h]%
	\begin{center}
		\begin{tabular}{ccc}
			\multicolumn{3}{c}{\includegraphics[width=9cm]{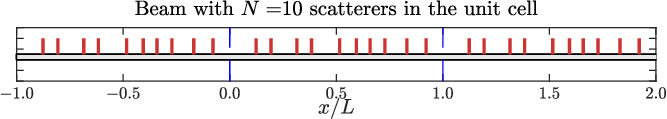}  		} \\ \\
			\includegraphics[width=5cm]{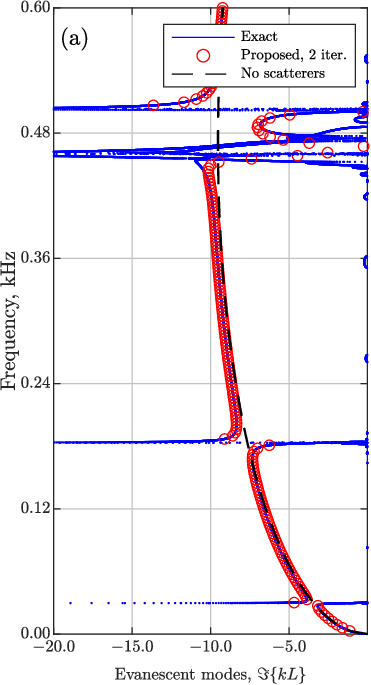} &
			\includegraphics[width=5cm]{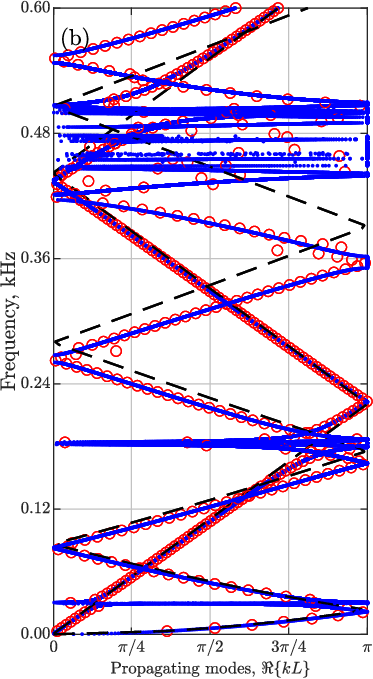} &
			\includegraphics[width=5cm]{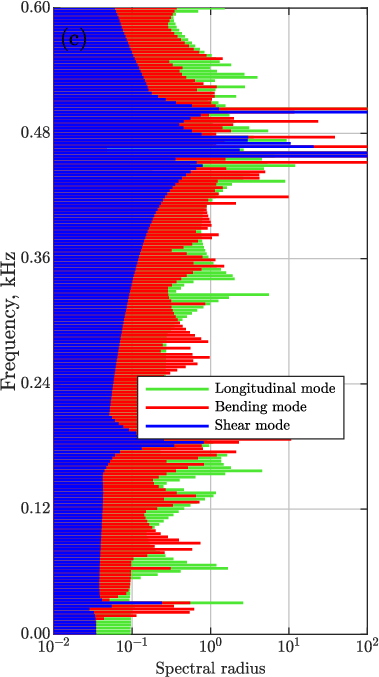} \\			
		\end{tabular}			
		\caption{(a) and (b) Imaginary and real part respectively of dispersion relation in Example 3, using the second-order iteration (Timoshenko beam with continuous resonators). Exact result (blue), Iterative approach with the second-order iteration (red markers) and unperturbed homogeneous solution (dashed-black). (c) Plot of the spectral radius of the Jacobian matrix $\mathbf{J}$, criterion of convergence is $\rho(\mathbf{J})<1$. In green, spectral radius associated to the longitudinal mode of the beam. In red, bending mode and in blue, the shear mode}%
		\label{fig08}%
	\end{center}
\end{figure}

Figure \ref{fig08} shows the dispersion diagram for the beam with the resonators. 
The two possible propagating modes can be distinguished: longitudinal modes (horizontal displacement, $u(x)$) and flexural modes (vertical displacement, $w(x)$).  The resonators can vibrate longitudinally or transversely. In the first case, vertical forces are induced, which modify the flexural modes of the medium. In the second case, horizontal forces and moments are induced in the beam, modifying both the longitudinal and flexural modes in the medium. Thus, several hybridizations in the flexural bands of the beam can be seen at frequencies $\omega / \omega_c = 0.03, \ 0.186, \ 0.46, \ 0.52$. In the regions around the resonances, bandgaps open with significant scattering. At other frequencies, the wave behavior reflects significant changes compared to the modes of the bare beam (represented by a black dashed line). Despite this, the proposed model reproduces the dispersion relation accurately, as shown in Figure \ref{fig08}, using 2 iterations, see Eq. \eqref{eq073}. The spectral radius of the Jacobian matrix $\mathbf{J}$ can be evaluated for each one of the propagating modes in Figure \ref{fig08}(c), where the regions reflect the measure of the perturbation of each mode and allow diagnosing the quality of the iterative procedure. The  exact results  used for reference have been obtained, as in the previous example, by the transfer matrix method. In this case since the kinematics of the waveguide is defined at each point by two components of displacements and a rotation, the transfer matrices have size $6 \times 6$. The code of the numerical model has been shared for interested readers in the Supplementary materials. \\

\begin{figure}[h]%
	\begin{center}
		\begin{tabular}{ccc}
			\multicolumn{2}{c}{\includegraphics[width=8cm]{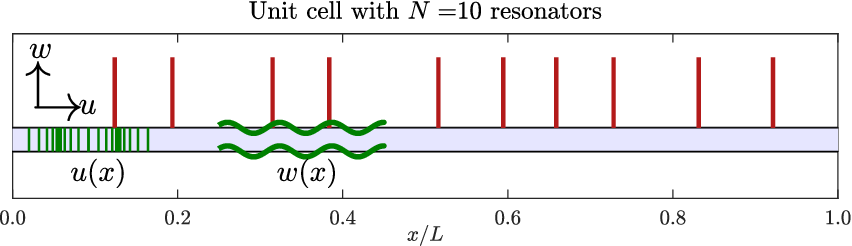}  		} \\ \\
			\includegraphics[width=8cm]{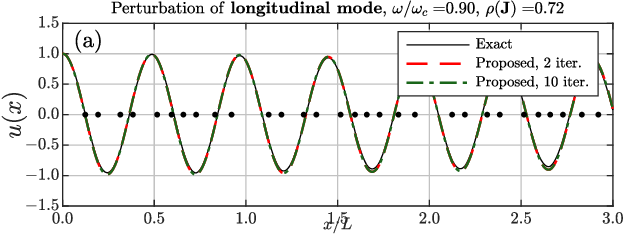} &
			\includegraphics[width=8cm]{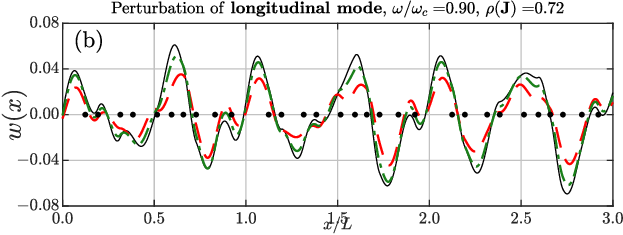} \\			
			\includegraphics[width=8cm]{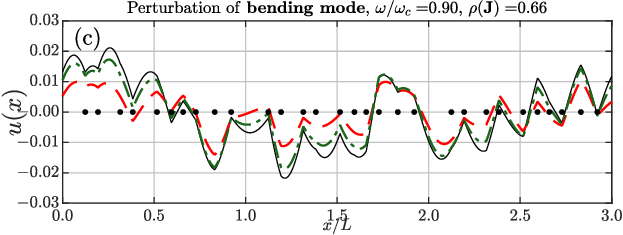} &
			\includegraphics[width=8cm]{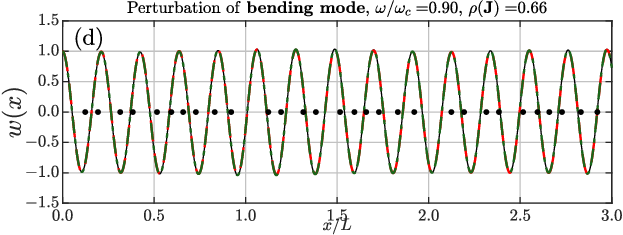} \\			
			\includegraphics[width=8cm]{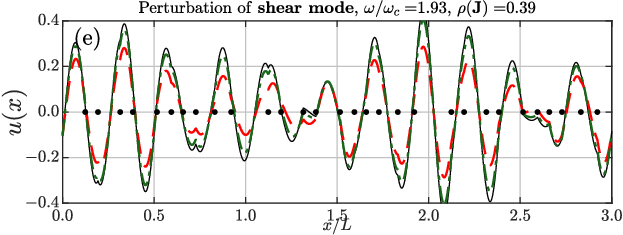} &
			\includegraphics[width=8cm]{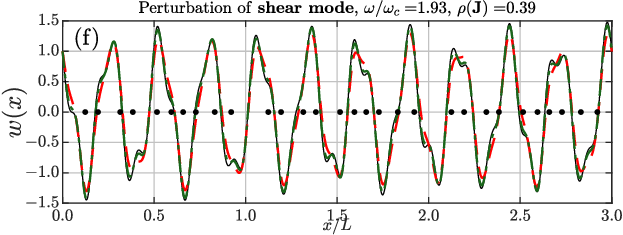} \\			
		\end{tabular}			
		\caption{Representation of propagating longitudinal, bending and shear modes for Example 3 obtained for a frequency of $\omega / \omega_c = 0.90$. Left plots, (a), (c) and (e) represent the horizontal displacement $u(x)$. Right plots, (b), (d) and (e) depicts the vertical displacement. In black, the exact mode, in red the proposed approximate method with two iterations, in blue with 10 iterations. }%
		\label{fig09}%
	\end{center}
\end{figure}

We will conclude this last example by showing the propagating modes for some specific frequencies. The results are shown in Fig. \ref{fig09}. It is worth noting that in this case, each mode includes both horizontal displacement $u(x)$ and vertical displacement $w(x)$. In the left column, Figures \ref{fig09}(a), \ref{fig09}(c), \ref{fig09}(e) show the horizontal displacement, while in the right column, Figures \ref{fig09}(b), \ref{fig09}(d), \ref{fig09}(f) show the vertical displacement. The modes obtained after 2 iterations, 10 iterations, and infinite iterations (exact mode case) are shown. Note that the three cases represented correspond to convergent simulations, as reflected by the spectral radius of each mode considered. The first two rows reproduce the longitudinal and flexural modes dispersed in the medium for the frequency $\omega = 0.90 \omega_c$. The longitudinal mode exhibits polarization in the transverse direction due to the presence of the resonators. The displacement $w(x)$ shown in Figure \ref{fig09}(b) is smooth, as the rotations of the cross-sections are continuous functions. However, the horizontal displacement $u(x)$ that arises in the flexural mode, Figure \ref{fig09}(c), has abrupt changes in the derivative. This is reflected by the model and reproduces the horizontal forces induced by the resonators. Note that the changes in slope coincide with the positions of the resonators, represented by black dots. The perturbation of a shear mode for the frequency $\omega = 1.93 \omega_c$ is shown in the last row, Figures \ref{fig09}(e) and \ref{fig09}(f). Despite the high value of the frequency, the model works and shows that the accuracy does not depend on the frequency but on the scattering characteristics of the model reflected by means of spectral radius $\rho(\mathbf{J})$, whose value in this case does not exceed unity, ensuring the convergence and feasibility of using the proposed approach.\\

To compute the exact waveform shown in Fig. \ref{fig09}, the transfer matrix method was employed. Initially, the wavefield at each resonator's location (i.e., at the cross-sections to the right and left of each $\xi_\alpha$) was determined. Subsequently, the wavefield between the resonators was filled using the modes of the bare beam, which resulted in the complete field along the unit cell at the frequency of interest, $\omega = 0.90 \omega_c$. In contrast, the approximate solutions after two and ten iterations were obtained using the recursive procedure described in Eqs. \eqref{eq054a} and \eqref{eq054b}, with the first and second iterations corresponding to the closed-form expressions in Eqs. \eqref{eq071} to \eqref{eq072b}. This iterative method is applicable at frequencies where $\rho(\mathbf{J}) < 1$, as defined in Eq. \eqref{eq064c}, and is generally computationally efficient and easy to implement. This is because each wavenumber $k^{(n)}$ and wavemode $\bm{\Psi}_n(\xi)$ can be computed using simple matrix products after evaluating the matrix $\mathbf{G}(k, \xi)$ from Eq. \eqref{eq045}. For the example presented here, with $N = 10$ resonators, up to 100 iterations can be completed in less than one second on a standard desktop computer. While convergence depends on the spectral radius $\rho(\mathbf{J})$, typically fewer than 10 iterations are sufficient to achieve accurate results. It is crucial to highlight that the primary advantage of the iterative method lies in its early iterations. Even the first or second approximation offers valuable insights into the dispersion behavior across the entire frequency range, particularly in regimes with weak scattering and for an arbitrary number of scatterers $N$ or large unit cells of size $L$. In such cases, taking the limit $N/L \to \infty$ in the first- and second-order expressions presents a promising approach for deriving analytical approximations of the dispersion relations. A more comprehensive investigation of this asymptotic regime is reserved for future work. \\

The proposed method has also been validated in phononic crystals with small inclusions, as presented in the Supplementary Materials (Numerical Example). This test case involves both longitudinal and flexural waves in waveguides containing narrow segments with altered material and geometric properties. The results demonstrate that the analytical expressions, even under moderate impedance contrast, accurately capture the dispersion relation outside the bandgap regions. In particular, the second-order iteration provides reliable predictions for both propagating and evanescent modes, provided the scattering remains weak, as quantified by the spectral radius criterion.

\section{Conclusions}

This study presents a robust analytical framework for deriving exact and approximate dispersion relations in generalized one-dimensional phononic crystals. By utilizing a unified mathematical model and iterative procedures, the methodology successfully accommodates various types of scatterers and resonators, demonstrating remarkable versatility across different waveguide configurations. The theoretical approach, validated through numerical examples, reveals its efficacy in capturing complex interactions in scenarios of weak scattering. Moreover, the iterative scheme offers significant advantages in terms of convergence and precision, providing accurate predictions of dispersion behavior while maintaining computational efficiency. These findings not only deepen our understanding of wave propagation in elastic media but also pave the way for designing advanced phononic structures with tailored wave control properties. Future work could extend this approach to higher-dimensional systems or explore the effects of stronger scattering regimes.

\appendix

\section*{Acknowledgments}

M.L. and V.R.-G. are grateful for the partial support under Grant No. PID2020-112759GB-I00 funded by MCIN/AEI/10.13039/501100011033, and also for the partial support under Grant No. PID2023-146237NB-I00 funded by MICIU/AEI/10.13039/501100011033. M.L. and V.R.-G. acknowledge support from 
Grants CIAICO/2022/052 and CIAICO/2024/318 of the ``Programa para la promoci\'on de la investigaci\'on cient\'ifica, el desarrollo tecnol\'ogico y la innovaci\'on en la Comunitat Valenciana'' funded by Generalitat Valenciana. M.L is grateful for support under the Grant UNI/551/2021: ``Programa de Recualificaci\'on del Sistema Universitario Espa\~nol para 2021-2023'', (funded by ``Instrumento Europeo de Recuperaci\'on (Next Generation EU) en el marco del Plan de Recuperaci\'on, Transformaci\'on y Resiliencia de Espa\~na'', a trav\'es del Ministerio de Universidades. R.W and R.V.C acknowledge financial support from the EU H2020FET-proactive project MetaVEH under grant agreement number 952039.

\section*{References}
\bibliographystyle{elsarticle-num} 
\bibliography{bibliography}





\end{document}